## Review Article

Brian Drummond*

# Understanding quantum mechanics: a review and synthesis in precise language

**Abstract:** This review, of the understanding of quantum mechanics, is broad in scope, and aims to reflect enough of the literature to be representative of the current state of the subject. To enhance clarity, the main findings are presented in the form of a coherent synthesis of the reviewed sources. The review highlights core characteristics of quantum mechanics. One is statistical balance in the collective response of an ensemble of identically prepared systems, to differing measurement types. Another is that states are mathematical terms prescribing probability aspects of future events, relating to an ensemble of systems, in various situations. These characteristics then yield helpful insights on entanglement, measurement, and widely-discussed experiments and analyses. The review concludes by considering how these insights are supported, illustrated and developed by some specific approaches to understanding quantum mechanics. The review uses non-mathematical language precisely (terms defined) and rigorously (consistent meanings), and uses only such language. A theory more descriptive of independent reality than is quantum mechanics may yet be possible. One step in the pursuit of such a theory is to reach greater consensus on how to understand quantum mechanics. This review aims to contribute to achieving that greater consensus, and so to that pursuit.

## Contents

*Corresponding Author: Brian Drummond:** Edinburgh, Scotland;
Email: drummond.work@phonecoop.coop

# 1 Introduction and context

## 1.1 Review subject: the challenge of understanding quantum mechanics

Understanding quantum mechanics is hard in six ways.

1. Quantum mechanics involves precise mathematical language and structure (the formalisms), but there is no consensus (a) on whether or not quantum mechanics must also involve interpretation; nor (b) on whether or not any such interpretation should lead to understanding. For some, the predictive power of quantum mechanics, allowing its use in practice, is all that matters. Others look either for a way of understanding the principles of quantum mechanics, or for what they might reveal about the world [1–9].

2. The controversy over interpretation is partly because many of our intuitions and concepts, and the non-mathematical language describing them, developed prior to the exploration of subatomic phenomena [10, 11]. In this sense, some familiar concepts and word meanings are *pre-quantum mechanical*, and might need to be modified [12, 13]. Quantum mechanics uses agreed mathematical language to analyze phenomena. There is, however, no agreement on a corresponding non-mathematical language to describe such phenomena [10, 14–19].

3. Some words used in writings on quantum mechanics can take a variety of meanings. Authors often do not make clear which meaning is intended. Any imprecision, in using non-mathematical language, can make understanding quantum mechanics harder than it needs to be.

4. There are many approaches to understanding quantum mechanics [20]. No approach has yet convinced the majority of physicists [21]. Many approaches highlight areas where further work is needed, if we are to reach greater consensus on how to understand quantum mechanics. For example, we might need to challenge pre-quantum mechanical concepts and intuition, or to use non-mathematical language with more precision and rigour.

5. The frequent failure to undertake a full analysis of a realistic measurement apparatus suggests that at least some consider it unnecessary and avoidable. Others undertake complex work to realistically model laboratory experiments [22, § 1.9.1], which can restrict the need for interpretative assumptions [23].

6. Underlying intellectual prejudices will very often affect how quantum mechanics is discussed [19][24, § 23.4][25, § 4.1][26–28]. Given the potential need to modify pre-quantum mechanical concepts and word meanings, it is important to consider what these prejudices might be, and to make assumptions explicit [11, § VI][29][30, pp. 16-18][31, pp. 6-8].

## 1.2 Review scope: understanding the theory rather than the phenomena

Work on the foundations of physics (whether by physicists or philosophers) shares some of the features of wider philosophy of physics [32, § 1(ii,iv)][33]. These include [29, § 2][32, § 1]: (a) going beyond the purely mathematical content of theories; (b) clarifying the meaning of central concepts; (c) examining conceptual ambiguities or inconsistencies; and (d) evaluating suitable ontologies. The scope of this review reflects these features.

What does it mean to understand quantum mechanics? There is no consensus on how to answer this question [34], but there are several useful approaches. One important distinction is between understanding phenomena and understanding theories [35][36, § 3.2]. On this point, some distinguish between (a) *explanatory* understanding of a *phenomenon* (relating the phenomenon to accepted items of knowledge) and (b) *pragmatic* understanding of a *theory* (being able to use the theory) [37, chs. 2, 4]. Pragmatic understanding of a theory is seen by some as necessary for explanatory understanding of a phenomenon [37, ch. 4].

One sufficient criterion for pragmatic understanding (of a theory) is an ability to recognize characteristic qualitative consequences of a theory without performing exact calculations. Such understanding depends on a person's capacities, knowledge and beliefs [38–40]. This approach to understanding appears relative (varying from person to person), but an objective approach is possible [41, § 4]. For example, understanding can be defined by reference to values and concepts shared widely among scientists (but need not necessarily coincide with truth or knowledge) [42–44]. Understanding involves explanatory relationships within a single theory [45], connecting theories through concepts which they have in common [46], and fitting theories into an overall framework or structure [47].

This review reflects these views of understanding.

It focuses on the understanding of quantum mechanics as a theory, rather than the understanding of the phenomena which gave rise to that theory. Indeed the review finds in the literature little, if any, agreement on how to understand the phenomena. In contrast, the review finds large areas of agreement on how to understand the theory. The review identifies qualitative characteristics of quantum mechanics, using concepts which are shared widely among scientists, and which quantum mechanics shares with other scientific theories.

## 1.3 Review aims: comprehensive coverage, precise non-mathematical language

There are many books and articles on various aspects of understanding quantum mechanics. Exploring this vast literature suggests the need for a new review with two specific features: (a) it should be comprehensive in scope, referring to enough of the literature to be representative of the current state of the discipline; and (b) it should be clear, concise and disciplined in its presentation and its use of language. This review aims to meet that need.

1. The review cites many sources, concentrating on work published in the last two decades. The main criterion used to select which sources to cite, is the degree of conceptual clarity contributed by a source. The main criteria used to select which insights to include are (a) the degree of agreement among authors who deal with the relevant point and (b) there being few, if any, authors offering convincing counter-arguments. The review generally reports on these criteria only by exception. Where the criteria are met, sources are cited without further comment. Where they are not met, alternative views are noted, and the extent of disagreement indicated. In general, the review is cautious: it reflects only what can be said with reasonable confidence.

2. This review concentrates on the conceptual aspects of quantum mechanics. To retain clarity in doing so, it aims to apply the precision and rigour of mathematics, to the use of non-mathematical language. The use of language in this review is disciplined: the review uses only non-mathematical language, so its internal coherence entirely depends on how precisely and rigorously it uses such language. The review's use of non-mathematical language is also precise (in that intended meanings of words are specified) and rigorous (in that such intended meanings are consistent throughout the review).

This review acknowledges that understanding quantum mechanics requires both (a) familiarity with the mathematical language of the formalisms, and (b) the use of non-mathematical language to relate these mathematical formalisms to the relevant physical phenomena. In the context of quantum mechanics, non-mathematical language is like a magnifying lens. The lens can make text clearer to some readers. Similarly, non-mathematical language can enhance understanding of the mathematics of quantum mechanics, for at least some users. Such language is not a substitute for the mathematics.

## 1.4 Review outline: a synthesised map: the understanding of quantum mechanics

To further enhance clarity, the main findings of the review are presented in the form of a coherent synthesis. The cited sources demonstrate that the significant elements of this non-mathematical language are consistent with the corresponding mathematics. Areas of disagreement, such as those outlined in Sect. 1.1 above, limit the extent to which the synthesis can be comprehensive: it is not always possible to synthesize apparently differing views.

The remainder of this Part 1 clarifies how this review and synthesis relate to wider discussions on the themes of reality, spacetime, probability and determinism. It aims only to explicitly clarify (a) the assumptions underlying this review and synthesis, and (b) the intended meanings of ambiguous terms. With that limited aim, brief reference is made to the literature to indicate that these assumptions and meanings are at least reasonable.

The next five parts set out the main findings of the review, using non-mathematical language in a precise way, and using only such language. Part 2 identifies statistical balance as a core characteristic of quantum mechanics. Part 3 highlights characteristics of quantum mechanical states. Part 4 uses these characteristics to review the issues of measurement, decoherence and uncertainty. Part 5 uses the insights from Parts 2, 3 and 4 to review some widely-discussed experiments, thought experiments and other analyses. Part 6 reviews some specific approaches to understanding quantum mechanics. It also notes how insights and perspectives from earlier parts are supported, illustrated, or developed by some such approaches.

The main findings of the review are summarized in Part 7. Part 8 is a glossary of intended meanings for many elements of the non-mathematical language used.

The review is, to the vast literature it reviews, what a map is to the territory it represents. It is necessarily concise and schematic. It can be read in isolation, but will be better appreciated when exploring the mapped territory. It is aimed at both (a) those who are new to the territory (and so need help to find their way around) and (b) those who know the territory well (for whom the map may highlight previously unnoticed aspects of, or connections within, the territory).

The glossary in Part 8 is as essential to this review and synthesis, as the legend is to a map. Different maps may use the same symbol to represent different features. In the same way, the intended meaning of elements of the non-mathematical language used in this review, may differ from the meanings intended by other authors, when they use the same words.

## 1.5 Moderate realism: physics might be able to describe independent reality

There are many shades of meaning within the concept of realism. Central to most of these is the concept of *mind-independent reality* [30, ch. 9][48, 49][50, ch. 2][51]. The phrase *mind-independent reality*, or simply *independent reality*, indicates a reality which exists other than only in human thought.

Idealism argues that mind-independent reality does not exist; in other words, only ideas in minds exist. A less extreme view is that of instrumentalism, which argues that science aims only for empirically adequate theories, and that notions, such as independent reality, which cannot be defined operationally have no scientific meaning,

Idealism and instrumentalism are not necessarily any more reasonable or scientific than the view that independent reality exists [22, § 2.3][30][52], and may in fact be less so.

- An independent reality might be the source of inter-subjective agreement (between observers) [53].
- Belief in an independent reality can motivate scientific pursuit and explain its progress [54, § 6].
- The structure in empirical data in quantum mechanics also seems to require explanation in terms of independent reality [55].

One use of the word *realism* is to express the view that the notion of independent reality is meaningful. This has been termed open realism [49], metaphysical realism [50, pp. 35–37][56, § 4.2.1], or ontic realism [22, § 2.3].

This open realism may be based on the idea that theories have to pass the test of experiment. It is hard to believe that experimental results reflect nothing other than processes and interactions within our minds [49]. Open realism is, however, careful to distinguish, in principle, be-tween (a) mind-independent reality (the world as it is) and (b) the phenomena (the world as perceived by our minds) [36, § 3.2]. This distinction results in open realism having two broad forms [49, § 4].

- One form considers that independent reality is unavailable for direct investigation (or veiled). On this view, physics only describes and analyzes phenomena. The extent to which physics might reveal features of independent reality is unknown [30, pp. 171, 174].
- Another form is known as physical, mathematical or scientific realism. It assumes that physics can aspire to describe, or in fact deals with, independent reality directly [30, part II][51, 57].

Whatever view is taken of realism in the broader context of science, or even physics in general, it is worth considering realism in a specific context [52, 58, 59]. Here the question is: can we understand quantum mechanics in the context of a wider, mind-independent reality (in the sense mentioned above)? Some suggest that the features of quantum mechanics are such that it may not be tenable to hold to physical realism [30][49, § 4][60]. This review finds that it is not yet fully clear that such a view is warranted. In the meantime, it appears to be worth pursuing the possibility that physics might be able to describe independent reality[54, 61][62, § 2.5][63–65].

This review contributes to that pursuit, which reflects a form of open realism which has been called *moderate realism* [66] or *pragmatist realism* [67].

## 1.6 Nonrelativistic quantum mechanics: time as independent from space

There are two major views on spacetime. Substantivalism holds that spacetime exists independently of matter. Relationism holds that facts about spacetime reduce to facts about matter: spacetime is only a human conceptual tool to describe phenomena [68–70]. These views can be combined with other approaches such as eternalism or presentism. In eternalism, or block universe, present, past, and future all equally exist in an unchanging spacetime. In presentism only the present is real [71]. Deciding on the best metaphysical approach to spacetime may depend on the specific context [72].

It is not clear which, if any, of these approaches is best suited to quantum mechanics [68, 71], in which the mathematical formalisms neither depend on, nor necessarily involve, the concept of spacetime [68]. The formalisms are used to analyze phenomena, taken to relate to physical systems, and these phenomena and systems are often described by reference to a spacetime background [71, 73].

In physics, spacetime is treated in three distinct ways: (i) in a nonrelativistic approach, time is treated as independent of space, (ii) special relativity treats time and space as interacting in spacetime, and (iii) general relativity treats spacetime as dynamically interacting with mass. Each of these three has a quantum counterpart.

1. It is sometimes appropriate to use nonrelativistic quantum mechanics, which treats time as independent of space. This quantum mechanics can be viewed as an approximation, valid in a limited regime, to a more fundamental theory [74, 75].
2. In other contexts phenomena and systems are treated in an integrated spacetime background. The main approach is relativistic quantum field theory, a group of effective theories which, again, can be thought of as approximate [76, 77].
3. General relativity has not yet been integrated with quantum theory. This might require one, or both, of them to be modified, or better understood [78–80]. For example, it might be that neither time nor space is continuous [79, 81–84].

Consideration of quantum field theory may shed light on some of the problems of understanding nonrelativistic quantum mechanics [74, 75, 85, 86], but has not yet resolved them. Indeed new foundational questions arise (see Sect. 5.8 below).

One test of the adequacy of any understanding of quantum mechanics, is the extent to which it is possible to apply and make sense of that understanding in quantum theory more widely, for example in quantum field theory [87]. This review focuses on (nonrelativistic) quantum mechanics. Its findings should, therefore, be taken as provisional in this respect.

In quantum mechanics the treatment of time is not fully resolved [88–96].

- Time is generally treated very differently from space.
- Time can either be measured by a clock external to the relevant systems, or it can be defined through the dynamical behaviour of those systems.
- Time can also be considered as an observable (although views differ on this [97–101]).

This review deals with nonrelativistic quantum mechanics in a way that is independent of which philosophical approach to spacetime is used as a basis for dealing with the relevant phenomena, and independent of the approach to time in nonrelativistic quantum mechanics.

## 1.7 Understanding probability: differing approaches, no consensus

The interpretation of probability in general is controversial. In general, probability analyzes uncertainty. Depending on which approach is taken to probability, the uncertainty is considered to be either (a) epistemic, an aspect of theories about nature (for example, reflecting ignorance), or (b) ontological, an aspect of nature itself (perhaps reflecting indeterminism).

Within each of these broad approaches, views differ on how to understand probability [102, § 8.2].

- The epistemic sense of probability can be viewed as one or more of subjective, logical (determined by the information available) [103, § 4], personalist (varying from person to person) [103, § 2], or pluralist (common to groups of people, or intersubjective) [104].
- The ontological sense can be viewed in terms of frequencies [105, 106], stochastic or deterministic dynamics [105, 107], the Humean mosaic (facts in the world) [105–107], propensities (single-case or long-run) [105, 106], or features of theories [106, 108].

Drawing clear distinctions between these differing approaches to, and views of, probability is, at best, challenging [104, 109, 110]. For example, Bayesian approaches, in general, treat probabilities as tools for making decisions based on incomplete information. Specific Bayesian approaches differ in emphasis: between subjective and objective [111], and among personalist [112, § III], logical [113] and frequentist [114]. It is unlikely that any single view of probability will apply in all contexts [115].

There are challenges within the relative frequency approach [109][111, § 4][115–117].

- Probabilities are defined by reference to the results of repeated experiments on a large number of identically-prepared systems.
- This seems to make the statistics of probability distributions purely empirical and objective but it depends on prior probabilistic assumptions: one cannot get a *probable* from an *is*.

The propensity approach was reintroduced precisely to deal with probabilities in quantum mechanics [118]. All propensity views attribute a disposition to unpredictable systems which is quantified by objective probabilities attributed to such systems. This disposition is viewed as an aspect of independent reality [106, 119]. Some views treat propensity as only a basis for an ontology, to be combined with another approach to probability, either objec-tive [115, 120], or epistemic [121]. It is not yet clear whether or not challenges arising in the propensity approach can be satisfactorily resolved [110, 122][123, § 5].

This review treats quantum mechanics in a way that is independent of the approach taken to understanding probability.

## 1.8 Using probability: a classical approach to quantum mechanical data

Probabilistic and statistical data in quantum mechanics can be fully analyzed in the framework of classical probability and statistics [124, § 8][125, 126].

- The probabilistic structure of data arising in classical physics theory represents a special case of classical probability theory, but
- the probabilistic structure of data in quantum mechanics represents a more general structure of classical probability theory.

There is inconsistency in how authors use the term "classical probability theory". This review follows those [124] who use "classical probability theory" to refer to the structure of probabilistic data common to quantum mechanics and classical physics. Others [125, 127] also highlight the structure of probabilistic data common to quantum mechanics and classical physics, but use "classical probability theory" to refer to the specific structure of probabilistic data unique to classical physics.

Kolmogorov's formulation of (classical) probability theory provides a rigorous base for calculations [117, 128]. It does not explain the physical nature of the probability measure [129]. Kolmogorov emphasised that no probability is unrelated to experimental context: each such context generates its own probability space [130–132]. This highlights that all probability statements are intrinsically (often implicitly) conditional [105, 115, 128, 130].

Probabilities in quantum mechanics can be analyzed in a Kolmogorov model as conditional probabilities, in any one of at least three ways:

1. by conditioning on detection [133]; or
2. by allowing for differences among runs of the same measurement [134]; or
3. by conditioning on experimental settings [130, 135].

Conditioning on experimental settings [135] acknowledges that probabilities in general reflect two elements of randomness: that of state preparation, and that of experimental settings. It reveals that probabilities in quantum mechanics reflect only randomness of state preparation. This might mean that any two random variables recorded under mutually exclusive conditions should, by default, use different sample spaces [131, 136].

## 1.9 Limits to determinism: not necessarily limits to understanding

In general, quantum mechanics does not predict the outcome of individual measurement events. At first sight this feature appears to be inconsistent with the so-called determinism of classical physics. This raises four questions.

1. What is meant by determinism?
2. Is determinism a feature of classical physics?
3. Is indeterminism a feature of quantum mechanics?
4. Is determinism inconsistent with free will?

There are no simple answers to these questions.

1. There is no single agreed definition of determinism [137]. This review will take determinism to imply that any possible group of systems (which are isolated from any other systems) will evolve in a single unique way from any possible initial conditions [138].
   Indeterminism is the absence of determinism.
   This review will classify each of determinism and indeterminsim as being either *ontological* (a feature of independent reality) or *epistemic* (a feature of our knowledge of independent reality).
2. Classical mechanics involves epistemic indeterminism. It is impossible to be certain of the precise position of a particle [139]. It is also impossible to be certain about any physical law [77]. Physics theories identify systems which can be treated as isolated, and laws describing how such systems evolve from initial conditions. Such theories have limited precision [77]:
   - no system can be fully isolated and,
   - for the known universe, the role of laws cannot be separated from that of initial conditions.

   Ontological determinism may, in principle, underlie epistemic indeterminism [140, § 4][141]. The basis for ontological determinism in classical mechanics can, however, also be challenged [77, § 5.5].

   - Applying a simple law can generate unpredictable behaviour [142], so determinism does not necessarily imply predictability [24, ch. 12][110][143].

   - There are also limits, in principle, within classical physics itself, to its analysis of situations to which it is commonly applied [137, 144, 145].
   - Moreover, even in classical physics there is also a limit, in principle, to the precision with which initial conditions can be specified [146, 147].

   It is therefore unclear what distinguishes ontological determinism. Classical physics, even in principle, involves uncertainty [139].

3. There is complete consensus that quantum mechanics involves, in *principle*, an indeterminism which is, at least, epistemic. Less widely acknowledged is its epistemic indeterminism in *practice*.

   - Quantum mechanical analysis requires the calculation of solutions to the Schrödinger equation.
   - For complex systems, this might not be possible [148] [149, § 6].

   Again, however, epistemic indeterminism is not necessarily inconsistent with some form of underlying ontological determinism [22, § 4.2.3][110, 150][151, § 5](Sects. 6.2, 6.3 and 6.7 below).

4. Establishing any relationship, between determinism and free will, depends on how each term is defined [137, 152–154], and on the context in which any potential link is considered [155].
   Fundamental ontological determinism would not necessarily conflict with practical epistemic free will (freedom to choose an initial state, regardless of its past, to check its future evolution) [156, § 3][157].

   - It is hard to argue for more than such epistemic free will, assuming that some form of law is in operation. Whether or not laws are deterministic or indeterministic, to modify our actions will modify our possible pasts (as even indeterministic laws fix objective probabilities) [158, § 5].
   - Apparently free choices are, therefore, in principle, linked to past events [157][159, § 3.8], although it is hard to see how, in practice, the existence of any such link could be established [160].
   - It is also true that the result of a deterministic evolution cannot necessarily be foreseen [138][159, § 5.5].

Thus it may be more reasonable to believe that free will is practical and epistemic only, rather than absolute and unconstrained.

There is ongoing debate on whether or not quantum mechanics is relevant to free will [156–159, 161].

This review treats quantum mechanics in a way that is (a) independent of whether independent reality is deterministic or indeterministic; and (b) independent of whether any indeterminism is ontological or epistemic.

# 2 A core quantum characteristic: prescribed statistical balance

## 2.1 Quantum mechanics: prescribing regularities among events

There is widespread agreement that quantum mechanics prescribes (specifies in advance) some aspects of expected future events relating to physical systems, in a range of possible situations [162][163, § 4.5]. Prescriptions (advance specifications) are made, collectively, as probability distributions. Understanding quantum mechanics, therefore, involves some of the challenges of understanding probability (see Sect. 1.7 above).

There is also widespread agreement that verifying the prescriptions of quantum mechanics is almost always statistical [22, § 4.2.3, § 6.4][164][165, ch. 9][166–168][169, pp. 206, 210][170, p. 99].

- Quantum mechanics does prescribe individual events in the limited sense of not ruling them out [171, § 6.2].
- It also sometimes prescribes probabilities of 1 or 0, which precisely prescribe individual events [172, § II.F][173, p. 20].
- In general, however, quantum mechanics prescribes only regularities among multiple events.

All the prescriptions of quantum mechanics are in terms of events. In this review event is taken to mean the instantiation of one or more properties within some region of spacetime [174, § 2]. The prominence of events in quantum mechanics has prompted some to explore the possibility that events, rather than systems, form the fundamental ontology of independent reality [2, 174][175, § 10.2][176].

- Such an approach does not necessarily rule out the usefulness of the concept of a system in understanding quantum mechanics.
- It does, however, highlight the possibility that such systems may be comprised of discrete events, rather than having a continuous existence.

- This review, in using the term system, intends to implicitly acknowledge that possibility. Where appropriate, the possibility will be explicitly highlighted.

That said, there are others who argue against an event based ontology, on the grounds both of its inconsistency with the corresponding formalisms and of its apparent dependence on assumptions about spacetime [177, § 2].

Among other events, quantum mechanics frequently prescribes the outcomes of system-apparatus interactions. For situations involving interaction with an apparatus in a measurement, the probability distributions can be verified using an ensemble of identically prepared systems [178, pp. 55-56][179, § 6.2, § 6.3].

- In such situations, the probability distribution characterizes measurement events collectively. In each such event, system and apparatus are combined [180].
- By analogy, the distribution characterizes spinning a specific coin with a specific spinning device [181, 182].

Thus, even if there are continually existing physical systems, quantum mechanics does not necessarily describe intrinsic features of such systems in isolation.

## 2.2 Statistical balance in quantum mechanics: prescribed, not explained

Some features of the prescriptions of quantum mechanics raise a significant challenge to understanding. For any given measurement type, in a series of measurement events, the outcomes (collectively) give statistics consistent with the prescribed probability. This feature is common in contexts to which probability is applied. The unusual feature of quantum mechanics is that, for some combinations of measurement types, the observed statistics indicate that the collective response of (what are taken to be) identically prepared systems to differing measurement types is not at all straightforward.

- This can be seen in an ensemble prepared so that the prescribed probability for a given outcome in a particular measurement type is 1 [183, § 8][184, § 2].
- Typically, empirical data for a second measurement type on that same ensemble are consistent with a prescribed probability other than 0 or 1 (differing outcomes in repeated runs of the second measurement).
- A claim that each member of the ensemble was, originally, such that a definite value could be attributed

to the second measured property, is inconsistent with the fact that the prescribed probability for a given outcome in the first measurement type is 1.

– This can be seen by subjecting each member of the prepared ensemble to a sequence of measurement events: type two and then type one [56, § 1.1.2].

– The type two measurement effectively prepares two new ensembles: in each, the prescribed probability for a given outcome in a type two measurement is 1.

– When the type one measurement is then made on each of these ensembles, the results are consistent with a prescribed probability other than 0 or 1, and so inconsistent with the prescribed probabilities for the originally prepared ensemble.

– On this basis, it appears that no definite value of the second property can be attributed to individual members of the originally prepared ensemble.

– It is, therefore, remarkable that the statistics of the collective outcomes of the type two measurement are balanced to be consistent with a prescribed probability which is neither 0 nor 1.

In this review, the phrase *statistical balance* refers to this intricately balanced collective response, to differing measurement types, as reflected in prescribed probabilities and empirical data [185, Part 3]. This statistical balance features even in the analysis of events treated as relating to systems which are single (no subsystems) and simple (no structure) [183, § 8][184, § 2][186, § 4]. It has been suggested that this is one of the fundamental features of quantum mechanics [184][185, Part 3][186], but it is seldom highlighted [184], except for widely-extended composite systems (see Sects. 3.6 and 5.2 to 5.6 below).

There seems to be a widespread, implicit acceptance that explaining this balance is not part of quantum mechanics [184, 186]. There are, however, some attempts to identify the source of this balance.

– The statistical balance may be seen as a new law of nature [185, Part 3].

– The balance may reflect some other theory underlying quantum mechanics (see Sect. 6.7 below).

– The balance may reflect a conservation principle which operates at the level of whole ensemble, rather than at the level of each ensemble member [187].

– Some approaches accept the balance as a feature of independent reality which requires changes to some pre-quantum mechanical concepts [123, § 6][188, § 8].

This review frequently refers to the statistical character of almost all prescriptions of quantum mechanics, and to the concept of statistical balance. In doing so, the review, like quantum mechanics itself,

– accepts the collective response, of an ensemble of (what are taken to be) identically prepared systems, to (separately or in combination) differing measurement types, is balanced, but

– remains silent on the explanation or source of this pervasive statistical balance.

# 3 Quantum mechanical states: characteristics and classification

## 3.1 States: core features and challenges

There is ongoing disagreement on how the concept of *state* in quantum mechanics should be interpreted [179, 189, 190]. There is, however, widespread (though not universal) agreement on the following two core features of the quantum mechanical state.

1. The word *state* refers to a mathematical term (the exact form depends on the formalism used) [191, 192].

2. The state allows inference of probability distributions for collective outcomes of future measurement processes [25, p. 65] (see Sect. 4 below): in practice it is such data that are compared with prescribed probabilities (see Sect. 2.1 above).

These two features raise four particular challenges.

1. What precise non-mathematical language should we use to refer to states?

   – The phrase *state of the system* is often used. This is, however confusing, because the state reflects knowledge about the *ensemble*, not about any individual system [178, pp. 56-57][193].

   – The phrase *knowledge about the ensemble*, however, itself raises further confusion. It implies that at least one other system exists, which somehow *knows* [162, § 7][172, § II.A].

   – Some suggest *preparation of the system* as a more appropriate term than *state* [30, p. 254][194, § 2.1]. This raises two difficulties: (a) a preparation process might not operate as intended [16][22, § 1.4.1][195, § 4], and (b) preparation is a challenging concept in cosmology [50, pp. 54–57][196, § 6.1].

- Some define a state as an equivalence class of preparation processes (a class of processes which produce ensembles of systems which cannot be distinguished by experiment) [8, p. 92][22, § 1.1][189, § 10.2][196, § 11.5][197, § 13.2].

2. How does the concept of state apply to closed systems?

   - In theory, the Schrödinger equation applies to such systems [198, § 1].
   - In practice, we cannot make external measurements on some closed systems, such as those which contain our solar system [172, § II.A][199].

3. How does the concept of state relate to spontaneous events (not triggered by measurement) [200–202]? Quantum mechanics should apply to spontaneous transitions like radioactive decay, passively recorded [202]. Quantum mechanics should also apply to unobserved transitions such as those in the earth's core, or in space [203, p. xiii].

4. Does a state always, necessarily, relate to an ensemble, rather than a single system [204, § 1][205, p. 228]?

   - It is, usually, not possible to determine an unknown state by investigating a single system. Such determination may, however, be possible in particular cases [173, pp. 20-23][206].
   - As noted in Sect. 2.1 above, quantum mechanics can prescribe some individual measurement events in a limited way and others rarely.

## 3.2 States: a comprehensive, synthesized, non-mathematical characterization

Taking all this into account, suggests the following careful, comprehensive, non-mathematical characterization of a quantum mechanical state, achieved by a synthesis of elements of several analyses [162, § 7][163, 164, 176, 193, 207] of the concept.

- In quantum mechanics, a state is a mathematical term containing the following probabilistic information relating to a physical system.

   - The state prescribes, generally in terms of probability distributions, aspects of expected future events relating to a statistical ensemble of such systems, in a range of possible situations.
   - These situations may include the systems remaining closed, and may also include the systems interacting with other systems.
   - Interactions may be with an apparatus (in a measurement process) or with an environment (even without such a process).
   - For measurement, the state prescribes, for each type, probability distributions (for outcomes of repeated measurement events of that type on an ensemble of systems) reflecting a statistical balance in collective outcomes, both within ensembles, and among ensembles for differing measurement types.
   - In this limited context, some states represent an equivalence class of preparations.

The above characterization is not limited by the concepts of measurement, preparation or knowledge. This characterization can therefore, in principle, apply to systems which are not observed. (Whether or not it is possible in practice to determine the state of such a system is a different matter.)

The above characterization refers to probabilities and ensembles but unusual cases can be accommodated by (a) noting that in some cases the relevant probability can equal 1, and (b) allowing in some cases for the ensemble to have only one physical member and many mental copies.

## 3.3 Using subensembles to distinguish pure states from mixed states

There is some variation among authors on the subject of pure states. Some reserve the term *state* only for pure states while others allow states to be either pure or mixed [204, § 1]. This review takes the term state to cover both pure states and mixed states, and distinguishes between the two in terms of subensembles of the relevant ensemble.

A pure state is a state for which the relevant ensemble is such that any subensemble of that ensemble is also represented by that same state [193][205, § 2]. Thus, if the pure state ensemble is divided into subensembles, expectation values of all dynamical variables for any subensemble equal those for the original ensemble, and those for all the other subensembles [169, p. 307][173, p. 8]. Loosely speak-ing, a pure state cannot be mathematically expressed as a simple aggregate of distinct states [197, p. 7].

A mixed state is one for which the relevant ensemble can be split in such a way that each subensemble is represented by a different pure state [193][205, § 2]. For a mixed state, there will be many ways in which the relevant ensemble can be split into subensembles represented by pure states. This suggests that although, again speaking loosely, a mixed state can be mathematically represented as a simple aggregate of pure states, such a representation does not convey physical meaning, unless further information is available [25, § 6.1][193][197, § 2.4][208]. A mixed state can, alternatively, be called a proper mixture [179, 189].

Depending on the formalism in use, a pure state can be represented by any one of at least three mathematical terms: a density matrix, a vector or a wavefunction [25, § 6.1.1][179, § 12.1][193]. In contrast, a mixed state can only be represented by a density matrix [179, § 12.2][193]. This review uses the term density matrix to refer to all states, whether pure or mixed, and does not use either of the terms vector or wavefunction.

## 3.4 States: no clarity yet on any simple relationship to independent reality

Sect. 3.1 above notes some aspects of the ongoing disagreement on the concept of state. This section outlines a further area of disagreement about states.

Is the state objective or subjective? Is the state ontological (an aspect of nature itself), or epistemic (an aspect of theories about nature)? The state prescribes probability distributions, so it is not surprising that making such distinctions unambiguous is challenging (see Sect. 1.7 above). What can be said is that assignment of a state reflects, at least, broad intersubjective agreement among agents, who are assigning a state to a given physical system, on the basis of a given set of data [103, 209]. This still allows two agents, with differing data about the same situation, to assign different states to the same physical system [23, fn. 29][162].

Despite the challenges and ambiguities involved, some authors have explored what link there might be between (a) the mathematical, quantum mechanical state, and (b) an ontic state, taken to be part of independent reality. Two broad groups of views have emerged, referred to (loosely) as the epistemic view and the ontic view.

There are two types of epistemic view [210]. Quantum mechanical states convey information about, or relate to, either (a) measurement, treated as a primitive, or (b) underlying physical (ontic) states. In the latter view, some-times called a mixed model, an ontic state may relate to more than one quantum mechanical state (each of which may encode probabilities for more than one ontic state) [211–214].

In the ontic view, sometimes called a segregated model, an ontic state relates to only one quantum mechanical state [212]. In the simplest segregated model, each quantum mechanical state fully specifies a single ontic state. In other segregated models, a single quantum mechanical state can relate to several ontic states.

It has been suggested, subject to assumptions, that only segregated model (ontic) theories reproduce the predictions of quantum mechanics [215]. Despite this, many still view the quantum mechanical state as wholly or partly epistemic [213, 216, 217]. Their view is tenable given the clear scope to reject or challenge one or more of the explicit or implicit assumptions [88, § 2.2][171, 214][218, § 4][219–222]. Ongoing arguments for the ontic view [223] appear to be similarly inconclusive [217, 224].

One challenge to any ontic view is how the mathematical term representing the quantum mechanical state relates to independent reality [192]. For example, there are many terms in a typical density matrix. It is not obvious how these many terms correspond to conventional four-dimensional spacetime [53, 225–228]. One suggestion is that an apparent four-dimensional spacetime emerges from a more fundamental realm with very many more dimensions [229]. The quantum mechanical state can, however, be directly linked to an ontic state in conventional spacetime [118, 230, 231]. For example, the many terms of the typical density matrix can be taken as coefficients of a multi-field. A field specifies properties, smoothly across spacetime, by reference to *each separate* point in spacetime. A multi-field specifies properties, again smoothly across spacetime, but by reference to *multiple* points in spacetime [230, 231].

An alternative classification distinguishes between the different parts of independent reality into which the mathematical, quantum mechanical state is mapped. In what is classed as an *empiricist* approach, the state is mapped into the macroscopic preparation and measurement apparatus (in a similar way to the first type of epistemic view noted above) [22, § 2.2]. The second type of epistemic view noted above, and the ontic view, are then together classed as *realist* approaches, where the mathematical state is mapped, in various ways, into microscopic ontic states [22, § 2.3]. Despite some clear benefits of the empiricist approach [22, § 2.4], the debate in recent years (between the epistemic view and the ontic view, as outlined above) shows that it has not yet achieved widespread support.

Overall, there do not yet appear to be clear grounds to accept any simple relationship between the quantum mechanical state and independent reality.

## 3.5 Superpositions: combining states, not necessarily combining systems

This section reviews a particular way in which quantum mechanical states can be mathematically combined to form new states. A combination of states does not necessarily imply any physical combination of systems.

There is some variation among authors on the subject of superpositions. Some reserve the term superposition only for pure states [197, § 2.2]. Others allow superposition states to be either pure or mixed [193]. This review takes the term superposition to cover only pure states, and defines superpositions by reference to the coefficients used in combining states to form the superposition.

As noted in Sect. 3.3 above, speaking in loose terms, a mixed state can be expressed as a simple aggregate of pure states. The aggregate is simple in the sense that all the coefficients in its mathematical representation are real, positive numbers [25, pp. 90-91][197, § 2.1][205, pp. 222-223]. In contrast, a superposition state is, in both technical and loose terms, a complex combination of different pure states. The combination is complex in the sense that all the coefficients in its mathematical representation are complex numbers [197, § 2.2][232, pp. 16-18]. Strictly speaking the real numbers are a subset of the complex numbers, and this may explain why some authors [193] take mixed states to be a subset of superposition states. In this review, however, the coefficients in the representation of a superposition are taken to be *non-real* complex numbers, and so the term superposition is restricted to combinations (of pure states) which are themselves pure states [197, § 2.2]. The superposition state has characteristics distinct from those of the two which combine [232, pp. 12-13].

Superposition states are pure [197, § 2.2]. As noted in Sect. 3.3, this means that any subensemble of the relevant ensemble is represented by that same state [193]. In particular, the relevant ensemble cannot be split into subensembles in such a way that each subensemble can be represented by one or other of the two states which were superposed. A superposition cannot be interpreted as suggesting that each system in the relevant ensemble could, before any measurement, be represented by one or other of the two superposed states [61, 233]. This can be demonstrated both in theory and by experiment [173, pp. 10-11][186, § 4].

A pure state cannot be mathematically expressed as a simple aggregate of two or more states (see Sect. 3.3 above), but it can be represented as a superposition of two or more pure states [123, § 4][232, p. 12]. For a pure state, there will be many ways in which that state can be mathematically represented as a superposition of other pure states. One or more of these different superposition representations of a pure state may be useful in the analysis of any given physical situation.

In common with all quantum mechanical states, a superposition state prescribes, in terms of probability distributions, aspects of expected future events relating to a statistical ensemble of systems, in various situations. As noted in Sect. 3.2, for measurement situations, the state prescribes probabilities for each type, which reflect a statistical balance in collective outcomes, among ensembles for differing measurement types. (The term statistical balance was introduced in Sect. 2.2 above.) Thus all states implicitly reflect the core feature of quantum mechanics: the collective response of identically prepared systems to differing measurement types is statistically balanced (see Sect. 2.2 above). In superposition states, however, this core feature is explicitly visible in the mathematical term representing the superposition state [2, § 3]. A density matrix for a superposition state contains, not only terms representing possible measurement event outcomes, but also extra terms representing the statistical balance between such outcomes. The terms representing possible outcomes can in principle be experienced in a single measurement event, but the additional terms can only be experienced as statistical balance between outcomes.

One demonstration of the type of statistical balance characteristic of a superposition state, with these additional terms, is the phenomenon of interference (see Sect. 5.1 below) [165, § 9.2][234]. For this reason, the additional terms are commonly known as interference terms, and their existence is referred to as coherence [234].

Nothing in the above outline necessarily implies or involves combinations of physical systems. The next section reviews how it applies to such combinations.

## 3.6 Entanglement: statistically balanced subsystem outcomes

When systems interact, or are considered together, any separate states for the systems are replaced by a new one for the composite system [183, § 15]. In theory, the composite system state could relate to an ensemble empirically identical to the combined ensembles relevant to any states for the combining systems. Such a composite system state may be pure or mixed. Typically, however, the composite system state is a superposition of composite system pure states, and so is an entangled state [25, p. 149]. In theory, the entangled state, in common with all superposition states, is pure (see Sect. 3.5 above). In practice, such purity may be short-lived, as explored further in Part 4 below (or may not arise at all). In the rest of this Part 3, entangled states will be taken to be pure.

A pure entangled state, in common with all pure states, is a state for which the relevant ensemble is such that any subensemble of that ensemble is also represented by that same entangled state (see Sect. 3.3 above). In particular, the entangled state relates to an ensemble empirically different from any combination of ensembles relevant to any states for the combining systems [189, § 7.2].

Like all quantum mechanical states, entangled states feature a statistical balance in collective outcomes, among ensembles for differing measurements on the ensemble to which they relate. (The term statistical balance was introduced in Sect. 2.2 above.) Three features are seen in the statistical balance of an entangled state.

1. The statistical balance of an entangled state includes statistical balance in collective outcomes, among ensembles for differing measurements on individual subsystems. (Like other statistical balances specific to superposition states (see Sect. 3.5 above), this subsystem statistical balance is reflected in specific terms (in the density matrix which represents the entangled state) which are called interference terms, and represents a form of coherence [235, § 2.3].)
2. This subsystem statistical balance includes balance between measurement event outcomes for measurements on widely-separated subsystems.
3. This subsystem statistical balance also includes statistical balance between outcomes for differing measurement types on differing subsystems.

It is the *combination* of the latter two of these features which leads to results which attract much attention. Links between experimental results for distant objects are not strange: such links arise in classical physics. Statistical balance in collective outcomes, for differing measurement types, arises for even a simple system [173, pp. 10-11][184][186, § 4]. The unique and defining feature of entanglement is statistical balance among collective outcomes, for differing measurement types, on far-apart subsystems [236]. This is considered further in Sects. 5.2 and 5.3 below.

The composite system entangled state provides probability distributions for each subsystem in the context of experiments spanning the wider composite system. If states are separately assigned to subsystems (in the context of experiments restricted to only one subsystem), then these subsystem states will prescribe probabilities which differ from those prescribed by the composite system entangled state (in the context of experiments spanning the wider composite system). In this limited sense, entanglement contrasts with the idea that composite systems can be explained in terms of their subsystems [25, p. 185][53, 237, 238]. The contrast is limited because, as noted above, the state is distinct from the system [239, 240]. Some analyses appear implicitly to reject this distinction [241].

## 3.7 Reduced density matrices: useful tools for limited purposes

From a composite system state (mixed or entangled), we can compute probability distributions, for measurement event outcomes, for experiments restricted to only one subsystem. This calculation uses a reduced density matrix for the subsystem. The reduced density matrix can be mathematically derived from the composite state [242].

The reduced density matrix is a coarse-graining of the quantum mechanical state for the composite system [243]. The reduced density matrix is sometimes called an improper mixture [179, 189], because it is mathematically similar to a proper mixture. (As noted in Sect. 3.3 above, a proper mixture is an alternative term for a mixed state.) The reduced density matrix is, however, (in relation to the wider composite system) not a quantum mechanical state at all: neither a proper mixture (mixed state) nor a pure state. An improper mixture, or reduced density matrix, can only be termed a state in the context of experiments restricted to only one subsystem.

This relates to the point, noted in Sect 3.4 above, that two agents, with differing data about the same situation, might assign different states to the same physical system. An agent with data limited to one subsystem can appropriately assign the reduced density matrix as a state for that subsystem. If however, the same agent (or a different agent) has data relating to the wider composite system, then only the composite system entangled state can appropriately be assigned.

The reduced density matrix can be used as a calculational tool, to give accurate probabilities for measurement event outcomes, for experiments restricted to the subsystem but will *not* give accurate probabilities when experiments include the wider composite system [242, 244–246]. As noted in Sect. 3.1 above, the Schrödinger equation applies to closed systems. Reduced density matrices relate to subsystems which, in the context of the wider composite system, by definition are not closed, and so, in that context, the Schrödinger equation will not apply [247].

# 4 Measurement, decoherence and uncertainty

One challenge to understanding measurement in quantum mechanics is that the word implies a division of the world into system, apparatus and measurer. This raises several questions [248]. How is such a split to be made? Can quantum mechanics apply to the apparatus (and to the measurer)? What if the system is closed, so that external measurement is not possible, as would be true in applying quantum mechanics to cosmology? This part considers these questions in six stages.

Sect. 4.1 outlines two limited accounts of measurement, which refer only to a system and an apparatus, and apply quantum mechanics to both. Sect. 4.2 considers the measurement problem. Sect. 4.3 reviews the extension of the quantum mechanical analysis to the wider environment of the measurement. Sect. 4.4 outlines two broad categories of decoherence theory, one of which allows quantum mechanical analysis of closed systems. Sect. 4.5 reviews to what extent decoherence might contribute to solving the conventional measurement problem, or explaining the approximate validity of classical equations of motion. Sect. 4.6 reviews the uncertainty relations and their implications.

Several alternative approaches have been proposed to gain information about systems other than by conventional quantum mechanics measurement. Such approaches are known as weak measurement [249, 250], protective measurement [224, 251] and interaction-free measurement [252, 253]. The understanding of such processes, and their results, depends on which approach is taken to understanding quantum mechanics generally and, within that, how conventional measurement is understood [206, 254–258]. These approaches are not considered further in this review.

## 4.1 Quantum mechanics can apply to both system and measuring apparatus

Von Neumann's approach included the apparatus in the quantum mechanical analysis, but led to an infinite regress (each time an apparatus is included in such analysis, a further apparatus, excluded from the analysis, is needed) [169]. This was mathematically expressed in a pro-jection postulate but the meaning was unclear [259, § 11.1]. The projection postulate rarely features in practical applications of quantum mechanics [260] and its interpretation as a physical process has been challenged as being untenable [22, § 1.6][233, 261].

In this review, the phrase *measurement event* refers to the interaction of a single member, S, of an ensemble of systems, with an apparatus, A. Both S and A have an essential role [248], and both should be analyzed by quantum mechanics. In this review, the word *measurement* refers to a series of repeated measurement events (single runs), on members, S, of a statistical ensemble of identically prepared systems, to explore a joint property of S and A [193].

The collective outcomes of the measurement events constitute the *result* of the measurement. One aspect of the pervasive statistical balance referred to in Sect. 2.2 above is the care needed in discussing the result of a measurement. The need for such care is stressed by several authors, as outlined in the following paragraph. Underlying the need for such care are the features of quantum mechanics noted in Sect. 2.1 above: in general it prescribes only regularities among multiple measurement events; it does not necessarily describe intrinsic features of physical systems to which it is applied; and it suggests the possibility that events, rather than systems, may form the fundamental ontology of independent reality.

The result of a measurement is not ascribed to the systems, nor to their preparation, nor to the measurement, but to the totality. The totality is a closed phenomenon, and the prescribed probability distributions refer to this totality [9][262, § 6]. Measured values do not necessarily exist beforehand [22, § 4.6.1] and are defined only for the particular combination of S and A [193]. If a property has not been measured, the formalism does not attribute any value [195, § 3]. Only one context justifies a claim that any member S of the ensemble was originally such that a well-defined value could be attributed to the property being explored in the measurement. That context is when all single runs give the same outcome [193].

## 4.2 Measurement appears to reveal a problem in some interpretations

There is no single "measurement problem". The term is used in various ways. It describes challenges that can arise in using quantum mechaics to analyze (some combination of) three groups of empirical phenomena.

1. Single runs of a measurement usually result in a single definite outcome [263, § 1].

2. Repetition of apparently identical single runs of a measurement can lead to different outcomes [263, § 2].

3. The result of a measurement can increase, to some extent, the ability to prescribe the results of further measurements on the same ensemble [263, § 3].

Whether or not these phenomena generate a problem depends on the approach taken to understanding quantum mechanics [22, § 3.1.2].

For example, if states prescribe collective results for measurement on ensembles, how does this relate (if at all) to the outcome of single runs [23, p. 327]? For those who limit the role of quantum mechanics to prescribing collective results for measurement on ensembles, this problem might not arise [163, ch. 3][209, § 3].

More specifically, how can a single observed outcome in any single run, be consistent with the final state for the composite system [S+A], arising from the Schrödinger equation [259, ch. 11]?

– This problem arises most clearly when the final composite state is a pure superposition state [259, p. 441].

– This problem also arises, however, even if the final state is a mixed state: in that case there is, again, more than one possible measurement event outcome. Consequently, this problem is not solved by proving that interference terms (see Sect. 3.5 above) vanish from a superposition state. Any solution must also show how a mixed state can be consistent with a single outcome [259, p. 443][263, § 1]. As noted in Sect. 3.3 above, a mixed state can be mathematically represented as an aggregate of pure states in many ways [264]. This makes it difficult to solve the problem [23, p. 338].

– Where this problem arises, it cannot be dismissed by arguing that, although the composite system [S+A] state is entangled and superposed, there is a state for one of the subsystems (for S or for A) which is neither. As noted in Sect. 3.7 above, for any entangled composite system state, considering either subsystem as represented by any state yields probabilities inconsistent with the composite system state [265].

Even more specifically, why is the particular outcome observed in a given single run, rather than another outcome [266, § 2.2.3]? This partly relates to statistical balance among outcomes: observation (in a single run) of one value rather than another, contributes to collective outcome statistics. As noted in Sect. 2.2 above, explaining statistical balance is generally seen as outside the scope of quantum mechanics, and so in this review, the term "measurement problem" will not include this question.

Most statements of the measurement problem assume an ability to solve the Schrödinger equation for every physical system. For a complex system such as an apparatus, however, it may be that the equation can, neither analytically nor numerically, be solved. If so, then quantum mechanics cannot be applied to the apparatus, and the measurement problem does not arise [148, 267].

Whether or not the problem can be solved depends on its premises and formulation in any given interpretation [164, 193][217, § 2.4][218][259, § 11.2][268, ch. 5][269, 270]. This is illustrated in the following section.

## 4.3 Including the environment in the analysis explains unique outcomes

Many analyses of measurement reflect an implicit assumption that the initial state of each of (separately) S and A is a pure state. While S can initially be in a pure (or a mixed) state, in practice it is inappropriate to assume an initial pure state for A [22, § 3.3.1][193].

Regardless of the initial states of S and A, the interaction between S and A will, at least in theory, lead to a final state representing the composite system [S+A] which is an entangled state. Including the environment, E, leads to the state representing the composite system [S+A+E] being an entangled state [164].

Before considering the effect of E, the state for [S+A] is an entangled state. As noted in Sect. 3.6 above, such states include both interference terms denoting statistical balance between subsystem outcomes (where S and A are subsystems), and terms representing the outcomes themselves.

Now considering the effect of E, within the entangled state for [S+A+E] a reduced density matrix can be derived for the subsystem [S+A], which will give accurate predictions for experiments restricted to [S+A] (as noted in Sect. 3.7 above). The reduced density matrix for [S+A], in common with the state for [S+A] before considering E, must have interference terms, in order to give accurate predictions, for experiments restricted to [S+A].

In practice, however, except for well-isolated, carefully-prepared systems, the interaction of [S+A] with E, leads to the interference terms in the reduced density matrix for [S+A] (representing statistical balance among subsystems S and A) becoming extremely small very swiftly [242]. Thus the reduced density matrix (for [S+A]) can be treated, approximately, as having no interference terms, similar to a mixed state density matrix.

The composite system [S+A+E] is still represented by an entangled state and so, in that wider context, no state can appropriately represent any individual subsystem. Measurement of an observable that only pertains to [S+A] cannot distinguish between the total ([S+A+E]) pure state and the (approximate, [S+A]) mixed state but, in principle, measurements involving E will always allow such a distinction to be made [242, 246, 247, 271]. This all follows from the nature of a reduced density matrix, as discussed in Sect. 3.7 above.

Quantum statistical mechanical analysis can build on this approach [193]. Such analysis can demonstrate that, after the interference terms in the reduced density matrix for [S+A] have become negligibly small, the remaining terms relax towards a thermal equilibrium [23, p.328], which is equivalent to a mixed state [25, p. 92].

At this stage it is possible to account for arbitrary subensembles of single runs. Large subensembles would statistically resemble the full ensemble, but more exceptional subensembles can act as a substitute for single systems. Information about any single run is gathered through all the subensembles in which it is embedded. This approach can explain the uniqueness of the outcome of each single run [50, § 9.6][193]. The framework, however, can neither describe nor explain why, for any single run, one particular outcome arises rather than another [272].

This use of quantum statistical mechanics combines: rigorous and detailed mathematical analysis of each element and stage of the measurement process; with careful use of approximations, such as disregarding events with very small probability, and ignoring possible recurrences after very long times [23]. Other analyses [259, § 11.4][270, § 7][273, 274] also suggest that this type of rigorously detailed analysis can explain the occurrence of just one outcome.

## 4.4 Types of decoherence: extending the quantum mechanical analysis

The questions noted at the start of Part 4 motivate an account of measurement which, in principle, allows quantum mechanics to apply beyond the measured system and apparatus, and also to closed systems. Decoherence theory can at least contribute to such an account.

As noted in Sect. 3.5 above, in the context of superposition states generally, coherence refers to the existence of interference terms, which represent only statistical balance between outcomes. Decoherence theories explore physical and mathematical processes which lead to the disappearance of such interference terms [246]. Such theories involve coarse-graining: the use of reasonable approximations [234, 275]. There are several approaches to the modelling of decoherence, which are being refined in the light of ongoing experimental testing [276].

There are two broad categories of decoherence theory.

1. Environment-induced, or extrinsic, decoherence involves the disappearance of interference terms, over time, induced by an external agent [234]. This involves mathematical analysis of the physical process of measuring a specific observable as described in Sect. 4.3. As noted in Sect. 3.6, in entangled superposition states, coherence includes statistical balance between measurement event outcomes for measurements on different subsystems. In the mathematical analysis, this balance is reflected in interference terms. As noted in Sect. 4.3, considering the environment initially introduces statistical balance in a higher order form [164] but, in practice, the interaction with E leads to interference terms in the reduced density matrix for [S+A] becoming extremely small very swiftly. A coarse-grained approximation, which ignores E and the residual balance, eliminates coherence [275]. Typically the environment analysed is limited to laboratory apparatus. Although, in principle, the environment could be extended to include the measurer, this is not possible in practice. As noted in Sect. 4.2 above, even for a complex system such as an apparatus, it may be that the Schrödinger equation can, neither analytically nor numerically, be solved. This would certainly be true of an environment which included a human (or feline) measurer.

2. Self-induced, or intrinsic, decoherence results from the basic properties of the system [234, 277]. Such decoherence is unrelated to measurement and so can apply to closed systems, for example in cosmology [277, 278]. Intrinsic decoherence is a relative process. In the mathematical analysis of a closed system, it is (notionally) split into an open subsystem (consisting of the parts, or aspects, of the closed system in which we are interested) and a residual subsystem (the closed system's other parts, or aspects). In a coarse-grained approximation, ignoring the residual subsystem, interference terms disappear (and so coherence is eliminated) for the open subsystem [275].

In this review the term intrinsic decoherence refers only to the coarse-graining approach described above. The same term is also used to refer to analyses which explore a possible breakdown of quantum mechanics [276]. Such analyses are quite distinct from the coarse-graining ap-proaches described above, and are not covered in this review.

Both types of coarse-graining decoherence can be linked within a common mathematical framework [275]. Extrinsic decoherence can be seen as a special form of the more general intrinsic decoherence [234].

## 4.5 Decoherence: not in itself a solution, but useful in particular contexts

Does decoherence offer a solution to the challenge of analyzing measurement in quantum mechanics? This question need a careful response [279, 280].

- Decoherence is a probabilistic concept and the only operational way of identifying decoherence is in the statistical behaviour of an ensemble [234].
- Decoherence theory can be applied in different interpretations of quantum mechanics [234]. As noted in Sect. 4.2 above, the nature and extent of the measurement problem depends on the interpretation. Thus the implications of decoherence theory for the measurement problem depend on specific interpretative framework used [245, 260, 266, 279, 281].
- Extrinsic decoherence theory involves a split between an ignored E and a considered [S+A], and so can be seen as involving something similar to von Neumann's infinite regress (see Sect. 4.1 above) [271].

Many believe that decoherence by itself does not solve the measurement problem [164][218, § 5][245, 247][268, p. 160][277, § 9][279, § 3.2][282]. This is because the reduced density matrix obtained by ignoring the environment is mathematically similar to a mixed state. A mixed state fails to explain, and is inconsistent with, the occurrence of just one outcome.

Some of the sources just cited, however, were published before the results of the quantum statistical mechanics approach, mentioned in Sect. 4.3 above [193]. That approach appears to explain the occurrence of just one outcome. It does so by a treatment of the measuring process which is more comprehensive than that of many papers dealing with decoherence [50, pp. 270–273]. The more recent papers refer neither to the quantum statistical mechanics approach [23, 193], nor to the other analyses mentioned in Sect. 4.3 above [259, § 11.4][270, § 7][273, 274], and so do not undermine the validity of such approaches.

More generally, a quantum-mechanical account of classical behaviour should also explain the approximate validity of classical equations of motion [283]. This has not yet been comprehensively done, and is likely to be complex [279, § 3.3][284], although some have already claimed success [285]. Decoherence may be part of the explanation but other components are likely to be needed too [280, 283]. For example, it may be that classical limit of quantum mechanics is classical statistical mechanics [286]. Explaining the motion of isolated systems may, in principle, not need to invoke decoherence, although it may be helpful in practice [164].

In summary:

- the combination of quantum statistical mechanics with extrinsic decoherence theory, allows the measurement problem to be solved, in those approaches to quantum mechanics in which it arises; and
- decoherence theory may yet form part of a successful strategy to explain, more generally, how classical mechanics emerges from quantum mechanics.

## 4.6 Uncertainty: a feature of statistics but not necessarily of systems

There are several different groups of uncertainty relations [194, 287–289][290, § 2].

- The Kennard-Weyl-Robertson uncertainty relations set, for two measurement types, a lower bound on the product of the standard deviations of measurement event outcomes, for an ensemble of systems.
- The Heisenberg noise (or error) disturbance uncertainty relations set a lower bound on the product of the noise (a measure of accuracy) in a position measurement and the resulting momentum disturbance.
- The Heisenberg joint measurement uncertainty relations set, for an apparatus jointly measuring A and B, a lower bound on the product of the noise in the A measurement and the noise in the B measurement.

All three groups of relations relate to measurement, but they do so in very different ways. For example, only the Heisenberg joint measurement group of relations deal with the extent to which two quantities can be measured at the same time. The differences between the groups of relations are not always sufficiently recognized. This can lead to significant confusion, mainly because the status of the three groups is very different.

- The Kennard-Weyl-Robertson uncertainty relations were derived as a rigorous mathematical consequence of the quantum formalism.
- In contrast, the precise terms of the latter two groups of relations have been challenged, and various ver-

sions have been derived [22, § 7.10][24, 25][175, § 9.2 - § 9.4][287–289, 291–293].

In the context of the Kennard-Weyl-Robertson relations, uncertainty is a precise statistical measure (the standard deviation) of the spread of future measurement event outcomes for large numbers of identically-prepared systems [292].

The Kennard-Weyl-Robertson relations can be read as a fundamental limitation on the possibility of preparing an ensemble of systems which, for any two measurement types, has statistical spreads that violate the inequality [22, § 1.7.1][25][173][178, p. 62][179, 194, 294].

The Kennard-Weyl-Robertson uncertainty relations depend on the state. Some states involve, for specific measurement types, no statistical fluctuation in measurement event outcomes, and the lower bound is zero [25, 289].

Heisenberg's semiclassical discussion of the noise disturbance relation (for a microscope) can, with care, be expressed in the quantum formalism [287]. Different versions can be derived [295]. Rigorous analysis of measurement interactions, direct computations and subsequent experiments, have all shown violation of at least one version of the noise disturbance relation [296–299].

The relations are often discussed informally in ways that suggest they relate to an individual system. There is, however, no obvious way to formally apply the relations to an individual system, in terms of values assigned to observables, or properties possessed [173, p. 14][175, § 9.2][185, Part 1].

- One reason is that the derivation and terms (such as standard deviation) of the relations are explicitly statistical.
- Another reason is that, as noted in Sect. 2.1 above, quantum mechanics in general prescribes only regularities among multiple measurement events, and does not necessarily describe intrinsic features of physical systems to which it is applied. On this basis the relations can be seen as highlighting further aspects of the statistical balance described in Sect. 2.2 above.

Thus, considering observables, and associated properties, relevant to the measurement types, the relations are silent on whether or not observables, or properties, might have definite values in a single system [25, 294, 300].

As noted in Sect. 1.6 above, the treatment of time in quantum mechanics is unresolved. Subject to being clear on the precise meaning given to time, and to precisely stating the aspect of energy considered, a range of time-energy uncertainty relations can be derived. The validity of any such relation will be subject to the terms on which it has been derived [90, 301][302, p. 46].

In summary, the uncertainty relations appear to specify features of the statistical balance among outcomes of measurement events in a statistical ensemble. There is no obvious way to formally apply the relations to an individual system, in terms of values assigned to observables, or properties possessed.

# 5 Experiments, thought experiments and other analyses

## 5.1 The two-slit experiment: no clear implications, but several possibilities

In the two-slit experiment, placing a quantum system on one side of the two slits produces a series of single, bright spots at specific, unpredictable locations on the screen on the other side of the slits. In this sense the system exhibits particle properties, at a specific time and place on the screen [303, 304].

If both slits are open then, when sufficiently many spots have appeared on the screen, an interference pattern emerges. At first sight, this resembles a classical wave effect, but on closer analysis the analogy is only partial [305].

- Any wave model would require the system to combine a wave aspect, from the slits to the screen, and a particle aspect at the screen [304].
- A wave model also suggests that two detectors, one behind each slit, should click simultaneously, but this generally does not happen [18][306, § 9, § 10].

Attempts to determine facts at either slit generally destroy the interference pattern [268, § 6.3][307]. The decision whether or not to make such a determination can be deferred until the system (if it is assumed to move from source to screen) would have passed the slits (a delayed-choice experiment). In such a delayed-choice experiment, the existence or non-existence of interference in the past, seems to be determined by a choice in the present [308]. Closer analysis of the relevant superposition states, however, reveals that the effect does not require present influence over the past [309].

The determination is often described as a which-way measurement, but this description is misleading [309–311]. As noted in Sect. 2.1 above, quantum mechanics suggests the possibility that events, rather than systems, may form the fundamental ontology of independent reality. Even if continuously existing systems are assumed, a detector in-teracting with a system does not necessarily imply that such a system has followed a particular route in physical space. Any such inference involves multiple untested assumptions about what happened prior to detection [128, § 3]. Recent experiments reveal possible evidence of aspects of a system's past, but there is not consensus on the meaning of the results [257, 312].

The results of the two-slit experiment are often summarized as single particle interference [232, pp. 8–9], which underlines the ongoing lack of clarity on how to understand any particle concept in quantum mechanics [22, § 2.4.4][313–317]. There are, however, several possible explanations for the difference between classical mechanics and the (verified) statistical prescriptions of quantum mechanics. One potential explanation is that interference may result from spacetime being discrete rather than continuous (see Sect. 1.6 above) [185, Part 1][318], or otherwise differing from that assumed by classical mechanics [319, § 3]. Explanations are also possible in some theories based on system positions (see Sect. 6.2 below). A third possibility is that interference may reflect the final state of each system combining with its initial state, in a time-symmetric approach [305] (see Sect. 6.4 below).

It has been suggested that the usual probability calculus breaks down in the context of quantum interference. This argument is misleading [136][268, § 6]. It is true that the data do not admit any simple Kolmogorov model which does not include rules for different contexts. The data require a Kolmogorov model with a probability space for each context (see Sect. 1.8 above). This combines data for distinct contexts: each context has a simple Kolmogorov model, but an interference term arises when adding probabilities from different contexts [185, Part 2][320].

Overall, even within the domain of appropriate inferences from the two-slit data to the behaviour of physical systems, the range of possibilities does not allow any clear conclusions on the nature of independent reality.

## 5.2 Einstein-Podolsky-Rosen: steering a reduced density matrix

A Kennard-Weyl-Robertson relation (Sect. 4.6 above) can be derived for position and momentum. One consequence of this relation is that, for any ensemble, there will be statistical fluctuation in outcomes for at least one of (a) repeated position measurement events and (b) repeated momentum measurement events. As noted in Sect. 4.1 above, only if all the single runs in a measurement process give the same outcome, can a well-defined value be attributed to the relevant property for the pre-measurement system. Thus the relevant Kennard-Weyl-Robertson relation implies that quantum mechanics is unable to attribute well-defined values to both position and momentum, for any given ensemble.

In the Einstein-Podolsky-Rosen paper [321], the authors analyzed a thought experiment, in which a pair of systems is created such that (a) if position (relative to the creation point) is measured for each, the results will be equal and opposite; and (b) if momentum for each is measured, the results will be equal and opposite. An observer, O2, might carry out a measurement, on system 2, of either position (and then predict with certainty the result of a position measurement on system 1), or momentum (and then predict with certainty the result of a momentum measurement on system 1).

In other words, the measurements on one system appear to affect the probability distributions for the other system, regardless of how far apart they are. The Einstein-Podolsky-Rosen argument was that either the properties of system 1 depend on O2's measurement result, or quantum mechanics is incomplete. Einstein-Podolsky-Rosen rejected the first possibility and were "thus forced to conclude" [321] that quantum mechanics is incomplete [25, § 9.3.1][53][159, § 3.2][268, § 7.1].

It appeared that O2 could choose which of two separate sets the state for system 1 will belong to: either the set of states for which, in a position measurement, each single run will have the same outcome; or the set of states for which, in a momentum measurement, each single run will have the same outcome. Schrödinger informally described this as O2 steering, or piloting, system 1 [322].

A more precise approach recognizes that the pair of systems must be treated as a composite system represented by an entangled state [159, § 3.2][212, 323]. As outlined in Sects. 3.6 and 3.7 above, one consequence is that any state which represents either subsystem in isolation will be inconsistent with the entangled composite system state, in relation to experiments on the whole composite system. The state for subsystem 1 (which O2 appears to steer) is a reduced density matrix for subsystem 1, in relation to the entangled composite system state.

A second consequence is that the collective outcomes of measurements performed on different subsystems are statistically balanced. As noted in Sect. 2.2 above, accepting the existence of such statistical balance does not imply any explanation of its source. Statistical balance does not necessarily imply causal connection, nor does it necessarily imply signalling between subsystems [324, § 5][325]. The statistical balance arises here because the pair, represented by an entangled state, is treated as a single entity in quantum mechanics. The Einstein-Podolsky-Rosen con-clusion is only "forced" if the pair is treated as appropriately represented by two separate states, one assigned to each of the pair [235, § 2.3][326].

## 5.3 Einstein-Podolsky-Rosen: statistical balance, not correlation

As noted in Sect. 3.6, the strangeness of entanglement is statistical balance among collective outcomes, for measurements of more than one observable, on distant subsystems. In the Einstein-Podolsky-Rosen context, each pair is a composite system with two subsystems. In theory, the outcome of a position measurement event for one subsystem is equal and opposite to the outcome of such an event for the other (although verifying this in practice is hard). Again in theory, the outcome of a momentum measurement event for one subsystem is equal and opposite to such an outcome for the other.

These features strongly resemble features of classical physics in which the term correlation might be used without controversy. It is therefore common to refer to correlations in this context, and to explore their implications in terms of causation [327]. There are, however, reasons to doubt that such an approach is appropriate. The word correlation carries at least some connotation of causation, or predetermination of properties unrelated to measurement. As noted in Sect. 4.1, however, in quantum mechanics, measured values do not necessarily exist beforehand, are specific to a particular system/apparatus combination, and can be attributed only after a measurement.

All that said, at first sight the combination of the following two facts still appears strange.

1. For a given ensemble of (Einstein-Podolsky-Rosen) pairs, quantum mechanics is able to attribute a well-defined relationship ("equal and opposite") between subsystem outcomes for each of position and (separately) momentum measurements.
2. For the same ensemble of pairs, however, quantum mechanics is unable to attribute to the composite system well-defined values for both (together) position and momentum.

The features of quantum mechanics highlighted in earlier sections of this review shed some light on the source of the apparent strangeness. As noted in Sect. 2.1, quantum mechanics does not necessarily describe physical systems, and usually prescribes only regularities among multiple measurement events. As noted in Sect. 3.2, quantum mechanics probability distributions reflect a statistical balance in collective outcomes, both within ensembles, and among ensembles for differing measurement types. As noted in Sect. 3.6, a composite system entangled state relates to an ensemble, empirically different from any combination of ensembles relevant to any states for the systems which combined to produce the composite. As noted in Sect. 4.1, if a property has not been measured, quantum mechanics does not permit any value to be attributed. Also as noted in Sect. 4.1, only if all the single runs in a measurement process give the same outcome, can a well-defined value be attributed to the relevant property for the pre-measurement system.

Overall, if anything is strange in the Einstein-Podolsky-Rosen context, it relates to some of the features in the immediately preceding paragraph, rather than to the first of the facts noted in the paragraph before that. For this reason, it appears more appropriate to use the term statistical balance, rather than correlation, to describe the phenomenon of entanglement. There is a sense in which entanglement is merely a more complex, and salutary, manifestation of the statistical balance that pervades quantum mechanics [184, 186] (being the core concept reflected in the quantum mechanical prescriptions for statistical ensembles: see Sect. 2.2 above). This idea (that entanglement reflects features which are also seen in quantum mechanical analysis of non-composite systems) was immediately noted by Schrödinger [183, § 10, § 11], and continues to be endorsed [22, § 6.4][328, § V.F].

## 5.4 Building on Einstein-Podolsky-Rosen: Bell explores the implications

Some counterfactual statements are true. For example, "in an Einstein-Podolsky-Rosen pair represented by a composite system entangled state, if both subsystems are subject to measurements of position, then the results will sum to zero". Care is needed in considering the implications of such a statement. As noted in Sects. 5.2 and 5.3 above, it does not necessarily imply pre-existing properties, causal connection, signalling or instruction sets that fix outcomes. Nor does it necessarily imply the truth of other counterfactual statements such as: "if the position of one is measured with result +1 and the position of the other is not measured, then if (counterfactually) the position of that other had been measured then the result would have been -1".

Some of these possible implications were explored by Bell [329]. Instead of position and momentum, Bell considered different components of a property known as spin. The relevant Kennard-Weyl-Robertson relation implies that quantum mechanics is unable to attribute well-defined values to both of two differing components of spin, for any given ensemble. Bell analyzed a composite system entangled state, which entails that the result of measuring any chosen spin component for one subsystem, can be predicted by first measuring the same component for the other.

Considering measurements on selected components of the spins, Bell hypothesized that if two measurements are made, when the subsystems are far apart, then the setting of the first measurement (as to which component of spin is to be measured) does not influence the result of the second. Bell suggested that this would imply that the result of the second measurement must be predetermined, which in turn implies the possibility of a more complete specification of the system than is given by the state.

Bell assumed such a specification, derived a resulting inequality, and showed that it is violated by the predictions of quantum mechanics. Bell concluded "In a theory in which parameters are added to quantum mechanics to determine the results of individual measurements, without changing the statistical predictions, there must be a mechanism whereby the setting of one measuring device can influence the reading of another instrument, however remote … , … instantaneously …" [329].

## 5.5 Bell inequalities: much exploration of assumptions but little consensus

Many papers written since Bell's 1964 paper have sought

- to make precise the assumptions on which the inequality depends,
- to explore what possible conclusions follow from these assumptions, and
- to establish to what extent such conclusions can be verified experimentally.

Little consensus has been reached in any of these areas. The rest of this section considers the assumptions on which this type of inequality depends. The next section considers what conclusions emerge, from the theoretical and experimental investigation of such inequalities.

Some analyze assumptions in terms of locality [175, § 8.6][293, 330], causality [175, § 8.7][331, 332] and local causality [333, 334]. Whether or not such assumptions necessarily imply counterfactual reasoning is not a straightforward question [335]. Others argue that Bell's inequality involves assumptions relating to distinguishability [336], determinism [158], ergodicity [337, 338], time-independent variables [339] or temporal locality [88, § 7].

Some focus on the assumption that changing an apparatus setting does not affect the distribution of any variables that determine the measurement event outcomes (measurement independence or free will). Supporters of this free will assumption argue that correlations between the systems and the settings chosen would have to be amazingly strong for it to be violated [161, 340]. This so-called conspiracy is, however, difficult to rule out [159, § 5.7.3][341][342, § 5][343][344, Appendix][345, § 4]. It is also consistent with the view that free will is only practical and epistemic (see Sect. 1.9 above). Arguments for the free will assumption may themselves involve circular reasoning [151, § 5.1].

Some question the apparent failure to correctly take into account the apparatus parameters, for different apparatus settings. Correctly treating the apparatus parameters, which amounts to assuming a form of contextuality, appears to prevent the inequalities from being derived [22, § 9.1.3][166, 346–352], but this view has been challenged [353]. An equivalent argument challenges the apparent assumption that there exists a single Kolmogorov probability space describing the statistical data collected by incompatible experiments [131, 135][175, § 8.5][181, 354, 355]. As mentioned at the end of Sect. 1.8 above, this may not be appropriate but, either way, none of the papers just cited refer to a previous claim that Bell's inequality can be proved without this assumption [293, pp. 83-85].

## 5.6 Bell inequalities: few clear implications from experimental investigation

Given the lack of consensus on what assumptions are, or should be, the basis for Bell-type inequalities, it is unsurprising that differing conclusions are drawn.

Some frame conclusions in terms of possible alternatives: realism or locality, but there is no single view of the meaning of these terms in such a juxtaposition [61, 306, 340, 356–359].

More fundamentally, some of those who challenge the appropriateness of mathematical or physical assumptions (as described in Sect. 5.5 above) deny that the inequalities indicate anything significant about either locality or realism [22, § 9.3.2][166, 342, 352]. Another fundamental concern is that many definitions of, or assumptions about, realism used in discussing Bell-type inequalities appear to be inconsistent with what are otherwise known to be core features of quantum theory [5, § 2.2][22, § 9.1.3][166, § 1][360].

This points to an even more fundamental reason to doubt that the inequalities indicate anything about either locality or realism [184, 186]. As noted at the end of Sect. 5.3 above, the entanglement analyzed in the inequalities is simply one manifestation of the statistical balance which pervades quantum mechanics. As noted in Sect. 2.2 above, few consider the explanation of that balance to be within the remit of quantum mechanics. In the absence of a theory to explain the pervasive statistical balance, the particular balance analyzed in the inequalities cannot justify conclusions on realism or locality. As also noted in Sect. 2.2 above, one explanation of the pervasive statistical balance is based on a conservation principle operating on average for ensemble. Such a principle can be used to derive the Bell inequalities without the need for any assumptions on locality or reality [361].

Overall, there is no consensus on whether or not quantum mechanics is local. Nor is there any consensus on which of the many possible meanings of locality is the most useful in this context.

Experimental investigation of the inequalities has also been ongoing. Experiments aiming to explore a Bell inequality involve multiple challenges. Many of these relate to the fact that, as noted in Sect. 2.1 above, both the predictions of quantum mechanics, and the data used to verify them, are in terms of events rather than systems. Some of the problems that can arise relate to unchallenged assumptions, incomplete analysis, insufficient statistics, incorrect statistical analyses, incomplete data (due to data discarding or postselection), spurious data (noise, dark counts, accidental counts) and corrupted data [362, 363]. For example, one common assumption is fair sampling, meaning that the observed outcomes of detections faithfully reproduce the outcome statistics of all emissions. Another assumption involves how best to pair together outcomes by reference to detection times.

No experiment can entirely overcome the multiple challenges. There is a widespread view that a series of recent experiments [168, 364–366] has dealt with all the significant challenges simultaneously. There is however ongoing and significant dissent from this view [367].

The fundamental issue remains, regardless of whether or not experiments succeed in overcoming the challenges. Bell experiment results are often claimed to constrain how quantum mechanics can be understood. Many such claims appear, however, to be inappropriate, given the lack of consensus (outlined above) on the theoretical aspects of the inequalities [132, 166, 338, 367–369].

## 5.7 The Bell-Kochen-Specker theorem: contextuality through mathematics?

The Bell-Kochen-Specker theorem is about the mathematical formalism of quantum mechanics. The theorem appears to show that quantum mechanics is inconsistent with the idea that measurement involves the ascertaining of a pre-existing value of a property [370]. Like the derivation of Bell's inequality, the Bell-Kochen-Specker theorem considers the possibility that parameters might be added to the formalism to determine individual measurement outcomes. The theorem shows that, in any theory involving such parameters which satisfies certain requirements, a contradiction arises. This implies that not all of the (mathematical) assumptions can be consistently held [25, 306].

As with Bell's inequality, there is a lack of consensus on how to understand these assumptions physically, and so a range of views on how various assumptions might be abandoned, and what implications would follow.

- In one view, systems have definite values for all magnitudes, but the result of a measurement depends on what else is measured at the same time. Thus an observable can only be fully defined by specifying the entire measurement context [293]. This does not rule out objective properties, but suggests that properties may not be knowable [370].
- In another approach, systems again have definite values for all magnitudes, but it is possible that some of those values cannot be revealed by measurements. In this case, the Bell-Kochen-Specker theorem does not imply that the results depend on the context [371].
- A third approach suggests that systems may not have definite values for all magnitudes, and so measurement event outcomes might not be determined in advance: a value-indefinite independent reality [372, 373]. In other words, systems may not have case-properties (determinately valued properties) for all their type-properties (determinable properties) [374], a radical revision to more common views [375].
- A fourth approach involves properties or systems being indistinguishable [376, § 4.1].
- A fifth approach takes potential states of affairs to be elements of independent reality [188].

Most approaches involve a form of contextuality [376, 377]. Contextuality is generally defined relative to one specific approach to quantum mechanics [371]. Broadly, contextuality suggests that the value assigned to a property depends on the measurement process [306]. Some suggest that experimental proofs of contextuality are possible but any experiment is likely to be based on one specific definition of contextuality [371], and contextuality is usually only sufficient, not necessary, to explain the relevant results [26]. The Bell-Kochen-Specker theorem itself cannot be directly tested experimentally (as any such test depends on how the assumptions are understood physically) [26, 326, 378], although this view has been challenged [379].

Overall, it is difficult to argue that the Bell-Kochen-Specker theorem necessarily supports any particular views on the interpretation of quantum mechanics [22, § 10.2.3].

## 5.8 Quantum field theory: no easier to understand than quantum mechanics

As noted in Sect. 1.6 above, when phenomena and systems need to be treated in the integrated spacetime of special relativity, the main approach is relativistic quantum field theory. Consideration of quantum field theory may shed light on some of the problems of understanding nonrelativistic quantum mechanics [74, 75, 85, 86]. Such consideration has, however, not yet resolved the problems of trying to understand quantum mechanics. Indeed new foundational questions arise [22, § 2.4.4][140, 143, 380][381, § 10.4.2][382]. This section outlines some of these new questions.

The fields in the formalism of quantum field theory are mathematical terms, which do not necessarily correspond to physical fields [177, § 2]. Like quantum mechanics (see Sect. 2.1 above), the predictions and verification of quantum field theory deal with events [143, 383], but the theory assumes the existence of entities as well as events. The standard model in physics involves particles and forces. To what extent do these notions reflect independent reality? There are significant challenges facing any such suggestion [56, § 6.4.2].

In quantum field theory, particles can be seen as aspects of the mathematical fields [140, 384, 385], and some argue that particles should be taken as ontological [177, § 2]. These particle aspects of fields are termed quanta, because they do not share all the features of particles in classical physics [386]. In quantum field theory, some terms in the mathematical analysis can be seen as reflecting creation and destruction of quanta [387, § 4], but such terms might not necessarily correspond to physical processes [22, § 2.4.4][140, 143, 388].

Similarly, some suggest that so-called "virtual particles" might not be physical [302]. Others, however, argue that virtual particles are no less part of independent reality than are quanta more generally [389].

How do quanta differ from classical particles? Unlike classical particles, quanta may not be capable of bearing labels (to allow them to be tracked over time) [384, 388] but this remains unresolved [17, 376]. Either way, could quanta still be particle-like because quanta are aggregable? In other words, can a determinate number of quanta be in a given system without the quanta having, in a formal sense, self-identity? Another way in which quanta may differ from classical particles is if there is not an appropriate sense in which they are localized [386].

Are quantum fields any less conceptually challenging than quanta? This is not yet clear [56, § 6.4.3]. Similar arguments to those that undermine the particle picture can also be turned against fields [390], but defenders of the field interpretation suggest that such arguments are not conclusive [391] [392, § 5].

One advantage of an empiricist approach (as outlined in Sect. 3.4) is that it avoids any choice between physical quantum fields and physical quanta. In this approach the quantum mechanical description corresponds directly to neither [22, § 2.4.4].

There are several approaches to the mathematical formulation of quantum field theory [56, § 6.3.5]. Which approach should form the basis of its interpretation? There is ongoing disagreement on this question [393, 394].

The ongoing experimental exploration of quantum field theory, helpfully and vividly illustrates two features which quantum field theory shares with quantum mechanics [193, pp. 7, 135]. One is the enormous size of the "apparatus" (such as the Large Hadron Collider in Geneva) relative to the "systems" under investigation. The other is the automatic recording and analysis of "measurement", without human intervention.

# 6 Specific approaches to quantum mechanics

Parts 2 to 5 above used only precise non-mathematical language to review characteristics of quantum mechanics which are central to understanding it. This Part 6 considers to what extent the findings of this review are supported, illustrated, or developed in a review (with this limited focus) of some historic and current specific approaches to understanding quantum mechanics.

As noted in Sect. 1.1 above, there have been, and currently are, many different approaches to understanding quantum mechanics. Some of these aim to develop a the-ory which might, in general rather than only rarely, describe individual systems (not just statistically prescribe collective outcomes) and so deal more fully with independent reality. Neither the Bell, nor the Bell-Kochen-Specker, analyses necessarily rule out the possibility of such theories, given the lack of consensus on their interpretation and implications (Sects. 5.5 to 5.7 above).

Reviews of the various approaches to quantum mechanics often start with the Copenhagen interpretation. The term *Copenhagen interpretation* was first used in 1955, by Heisenberg [395, § 4], and implies some combination of the views of different physicists associated with, or influenced by, Bohr [396, p. 462]. Some have attempted to specify core elements of this interpretation [8, § 3.4][22, § 4.1][397, § 3.3]. These analyses acknowledge, however, that there is no single, agreed definition of the Copenhagen approach [292, 395]. For this reason, Sect. 6.1 outlines only the approach advocated by Bohr. Some views outlined in later sections (for example, some in Sect. 6.6) have also been identified as Copenhagen interpretations [396, § 3].

## 6.1 Bohr: is a theory more descriptive than quantum mechanics possible?

Bohr's own approach was often based more on intuition and philosophy than on mathematical analysis [398]. This is partly consistent with the approach to understanding quantum mechanics outlined in Sect. 1.2 above: the recognition of qualitative characteristics, through concepts shared with other scientific theories. Bohr, however, did not use non-mathematical language in a disciplined, precise or rigorous way. For example, what Bohr meant by "complementarity" is the subject of ongoing debate [22, § 4.6.3][399–403].

Three aspects of Bohr's approach were, however, clear.

One is that, for Bohr, the formalism applies only to phenomena ("observations obtained under specified circumstances") [404, p. 64][405, § 3.3]. A phenomenon involves a system interacting with an apparatus, leading to an outcome. Consistent with the outline in Sect. 4.1 above, this is distinct from the system in isolation [406]. It appears that quantum mechanics, for Bohr, does not deal with systems, but with events [292]. This highlights that, as noted in Sect. 2.1 above, quantum mechanics does not necessarily describe the physical systems to which it is applied, and usually prescribes only regularities among multiple events.

A second clear aspect was Bohr highlighting that, in quantum mechanics, if one experimental arrangement permits the unambiguous use of the concept of position, then a different arrangement is needed to permit such use of the concept of momentum. For Bohr, a spacetime description meant determining the position of a system, but causality was linked to the conservation of momentum. Bohr characterized these two modes of description, spacetime and causal, as complementary but mutually exclusive [22, § 4.6.3][398, 407].

Thirdly, Bohr went on to suggest that any analysis of phenomena combining such complementary descriptions, is "in principle" excluded [408]. While this was clearly Bohr's view, both its meaning and its basis were unclear. It may be that Bohr's philosophy rules out such analysis [396, § 2(b)], but a theory which is more descriptive of physical systems in isolation than is quantum mechanics remains a possibility for those whose philosophy differs from Bohr's. In a similar way, a theory more descriptive of physical systems in isolation than is quantum mechanics remains a possibility for those [409] whose assumptions differ from those who deny such a possibility [410].

Another possible understanding is that Bohr was "in principle" excluding only the possibility that a value could be attributed to each quantum mechanical observable for the system independently of measurement [22, § 4.2.2]. This is not necessarily inconsistent with the possibility of a more fundamental theory underlying quantum mechanics, and there is evidence that Bohr was open to that possibility [22, § 4.2.2].

Some form of such analysis is certainly possible [22, ch. 7][411]. Mathematically, it appears that in quantum mechanics, any such analysis cannot be exact (in the sense of avoiding approximations) [412]. Even accepting this, neither logic, nor experiment, nor formalism, demand exclusion "in principle", even in a future theory, of a precise, objective, ontic description of independent reality. Experimental facts concern phenomena (system-apparatus interactions), not isolated systems. Complementarity of contexts (of phenomena) does not imply complementarity of properties (of systems) [185, Part 1]. Ontology may not be observable [381, p. 25][413].

Thus a theory more descriptive of physical systems in isolation than is quantum mechanics remains a possibility.

## 6.2 De Broglie and Bohmian theories: more descriptive but no less peculiar

So-called "hidden variable" theories were among the earliest attempts to develop a theory more descriptive than quantum mechanics. In such theories quantum mechanical analysis prescribes only some aspects of future events relating to the system, and the values of hidden variables provide a physical description of systems in isolation [25]. The term hidden is potentially misleading: the values of hidden variables are determinable in principle [53] and in practice [414], but are "hidden" from quantum mechanical prescription or control [22, § 10.3.4][151, § 3.3].

De Broglie's 1927 pilot wave theory had distinctly nonclassical dynamics. Bohm then revived these dynamics in a form which suggested to some that de Broglie-Bohm theories reflected classical principles [268, ch. 11][415]. Bohm, however, used the term quantum nonmechanics [387, § 2][416], emphasizing the non-classical nature of the dynamics [219, 417].

Bohmian theories aim to describe a situation as it exists independently of observation [416], and so to allow physics to describe mind-independent reality (see Sect. 1.5 above). This view of independent reality, however, appears to be as hard to understand as is conventional quantum mechanics [25, p. 160][387, § 2][417][418, p. 15][419].

For Bohm, a measurement event outcome was determined by hidden parameters for both apparatus and system [420, § 5][421, § 5]. Bohm agreed with Bohr on the fundamental role of the measuring apparatus as an inseparable part of the observed system (Sect. 6.1 above). Bohm differed from Bohr, however, in allowing the role of the apparatus to be analyzed, in principle, in a precise way [56, § 5.3][420, § 9], consistent with the outline in Sect. 4.1 above. Bohm's theory thus differs from the type ruled out by von Neumann's earlier theorem, in which measurement event outcomes depend only on the state for the system [420, § 9]. There is ongoing debate on the merit or otherwise of von Neumann's earlier theorem [422, 423].

There is a range of approaches to Bohmian theories [424] but most of them share the following five features.

1. The full description of a system combines the state and the configuration of the system in three-dimensional space [425].
2. Individual systems possess a definite position and their subsequent positions are determined by a quantum potential or a velocity field. Their positions are also guided by the state, which itself evolves according to the Schrödinger equation, and there is compatibility between this effect, and that of the potential [425]. Individual systems do not necessarily possess any properties other than position [50, pp. 177–179][419, 426]. Similarly, systems do not necessarily follow well-defined trajectories [200, § 6] [254, 418]. Bohm suggested that what appears to be a system might be series of events, so close one to the other that they look like a continuously existing system [387, § 2], in line with the possibility noted in Sect. 2.1 above.

3. The main state is a composite system state, notionally assigned to the known universe. The concept of a state of the universe raises significant practical and conceptual difficulties [427, 428]. These difficulties have limited effect on the application of Bohmian theories, however, because only conditional or effective states are associated with the systems under analysis [219, 239, 428, 429]. Referring to Sect. 3.7 above, relative to the notional composite system state, the effective state corresponds to a reduced density matrix for the (sub)system under analysis [430, § 3]. An important assumption in these theories is that, at some time, there is a random distribution of the positions of the systems under analysis, with respect to the relevant composite state. This assumption is known as the distribution postulate or the quantum equilibrium hypothesis [56, § 5.1.2][414, 425].

4. The guiding field in mathematical space has consequences in physical space [431–433]. An example is the Bohmian explanation of the two-slit experiment (Sect. 5.1) [268, § 6.1.1][434]. In some versions, the state (or guiding field) features in a physical law, rather than as part of independent reality [239, 428], although making this distinction is not necessarily straightforward [219]. In other versions, the state reflects properties of the systems [435]. In a third approach, both systems and state are ontological [228]. This raises the challenges common to all attempts to link quantum mechanical states to ontic states, as referred to at the end of Sect. 3.4 above. In a Bohmian context, however, there are more grounds for the ontological state to exist in conventional spacetime (rather than in a much higher dimensional realm), for example as a physical multi-field [230, 231].

5. Bohmian theories generally make only statistical predictions and so empirically verifying these prescriptions is almost always statistical, in the same way as it is for other approaches to quantum mechanics (see Sect. 2.1 above). The conceptually different feature of Bohmian theories is that the statistical character of the predictions is accepted as be-ing attributable to a lack of knowledge [50, § 5.6][56, § 5.1.2][436]. That said, there is not consensus on how probabilities are to be interpreted in this context [437]. The predictions of Bohmian theories for the collective outcomes of any experiment are generally held to agree with all unambiguous quantum mechanics predictions [50, § 5.1][414, 425]. This view has been defended against various challenges [438], although these continue to arise [439, 440]. There is ongoing experimental work to explore how modifying the assumptions might lead to results which deviate from quantum mechanics [421]. Consideration of measurement in the context of Bohmian theories requires care for two reasons. (a) As noted above, measurement outcomes depend on both the apparatus and the system and so the result of a position measurement will in general not reveal the position of the system prior to measurement [22, § 10.3.4]. (b) A full measurement analysis of a composite [S + A], as outlined in Sect. 4.3 above, in terms of any Bohmian theory would be so complicated that, in practice, it might be impossible [193].

The Bohmian approach illustrates that it is possible to develop a theory more descriptive of physical systems in isolation than is quantum mechanics. It also provides a formalism which allows classical hydrodynamic techniques to be used in quantum mechanical analysis, regardless of whether or not that formalism is understood in terms of a Bohmian theory with the above characteristics [441].

## 6.3 Relative state interpretations: applying the formalism to closed systems

The relative state interpretation was developed in response to the need (outlined in Sect. 3.1 above) to apply quantum mechanics to closed systems [442].

In this approach [442], any system that is subject to external observation is treated as part of a larger (composite) isolated (unobserved) system. Any state assigned to a subsystem of this composite system corresponds to a unique relative state for the remainder of the composite system. In a measurement, the apparatus, A, which interacts with system, S, is a subsystem in the larger isolated [S+A] system. In each single run of measurement, the A state branches into several different states. Each branch represents a different measurement event outcome, and the corresponding relative state for S. All branches exist simultaneously; Everett's original paper does not directly link branches to independent reality, partly due to the challenge of finding appropriate non-mathematical language [443].

The a quantum mechanical state is mathematical (Sect. 3.1 above). Assuming that a system is an aspect of independent reality, a relative state interpretation must, therefore, supplement the formalism in some way, or assume that the mathematical state describes physical facts [1, 171, 444].

– Most relative state interpretations involve some combination of explicitly adding worlds to the formalism, or invoking intrinsic properties of the mind, or using decoherence theory (Sect. 4.4 above) [56, § 5.2][445].
– A combination of the first and third of these approaches may be consistent with determinism [446].
– In other suggested approaches, branches are factual and counterfactual descriptions of one world [447], or differing trajectories, in spacetime, of point-like elements with a local internal memory [448].

Relative state interpretations take the terms in the mathematical state to be branches, and so to be outcomes. There are many ways to expand a composite system state (whether pure or mixed) as a combination of other states (see Sects. 3.3 and 3.5 above). Everett's approach focused, for a given observable, on an expansion such that, for each term in the expansion, the observable has a definite value. There are, however, possible expansions in which the observable has, for any term in the expansion, an indefinite value. How does the relative state approach link such expansions to measurement events? For some, no link is needed [447, § 4]. Others invoke decoherence in this context [260, 449, 450]. It might seem inappropriate to apply extrinsic decoherence (based on the effect on [S+A] of a wider environment E) in a relative state approach (which treats [S+A] as a closed system) [451], but the use of intrinsic decoherence seems legitimate (see Sect. 4.4 above). The use of decoherence in this context has, however, been challenged on other grounds [452].

Understanding probability is a particular challenge in relative state interpretations, regardless of which approach to probability (see Sect. 1.7 above) is taken.

– How can a measurement event outcome be uncertain, if all the components (of the expansion relevant for that measurement type) exist? One suggested approach is that, in the context of a world which branches, an agent is uncertain about which branch that agent will be in after branching. Alternatively, in the context of a multiverse of possible worlds (none of which branch), an agent is uncer-tain about which world that agent is in [453]. The concept of branching time (or spacetime) may be relevant [454, 455].

   – Why is that uncertainty quantified according to the formalism? There is disagreement over whether or not the quantum mechanics probability rules can be derived (in this context) in a decision theoretic approach [444, 456–459].
   – Can probabilities be empirically confirmed? The coherence of empirical verification is also unresolved [161, 459–461][462, § 4].

Relative state approaches illustrate challenges which arise through treating the mathematical formalism as directly relating to ontic states of systems [170, § 2]. These issues are in addition to those outlined in Sect. 3.4 above.

## 6.4  Time-symmetric approaches: is this challenging concept helpful?

Is there a substantial difference between the two temporal directions: towards past and towards future [463]? There are logical, epistemological and general scientific objections to any such difference [464], with differing views on their validity [465–467]. In the specific context of quantum mechanics, features such as those explored in Sects. 5.2 to 5.6 above, suggest some combination of time symmetry [468, 469], reverse causality [88, § 5.2][470–472], an adynamical spacetime [222], or a three dimensional timeless space [473].

In the context of an already ambiguous role of time in quantum mechanics (Sect. 1.6 above), these possibilities raise significant challenges for understanding (in the sense outlined in Sect. 1.2 above). These challenges are illustrated by three time-symmetric approaches: the transactional interpretation, the consistent histories approaches, and the two-time, two-state approach.

   1. The transactional interpretation aims to provide an observer-free account of measurement [172]. It assumes that an emitter sends an offer wave, possible absorbers each receive part of it, and send confirmation waves, backwards in time, to the emitter, which chooses one of them as the initial basis for a transaction. Repeated emitter-absorber wave exchanges, in both time directions, develop a spacetime standing wave to complete the transaction.
   This fits well with some features of the mathematical formalism, but raises significant conceptual challenges. For example, the nature of the waves is either "somewhat ephemeral" in ordinary space [474], or ontological in an "extraspatiotemporal domain of quantum possibility." [123, § 6][475]. Despite such challenges, there are ongoing suggestions that this approach makes a positive contribution to understanding quantum mechanics [200, § 4][253][387, § 3][476].

   2. The consistent histories formalism only assigns probabilities to consistent sets of histories [477]. A set of histories, or sequences of measurement events, is consistent if there is a single probability space accommodating them all. Such sample spaces are referred to as frameworks [198, 478].
   True and false are understood relative to a framework [479]. This in itself is conceptually challenging. It requires accepting that there is not a unique, universally true state of affairs. Reasoning must be done in a single framework: incompatible frameworks must not be combined. All frameworks are equally valid, varying only in their usefulness [328], but some propose restricting the valid frameworks [480], or amending the formalism [477].
   The histories approach claims to be compatible with the idea of independent reality [203], but is criticized as adding little description of that reality [477, 481, 482].

   3. The two-time, two-state approach limits the ontology of a system to those properties which, in a measurement event, have a definite outcome [305].
   Each system is represented by two quantum mechanical states, one evolving forwards in time, the other backwards in time, in a two-state formalism which has been separately developed [483].
   Measurement event outcomes reflect the combination of the post-measurement state (evolving backwards in time) and the pre-measurement state (evolving forwards in time). This analysis implies effects which might underlie apparent paradoxes in quantum mechanics [484, 485]. To the extent that this explains measurement event outcomes by knowing what those outcomes are, it contributes little to understanding quantum mechanics [486, § 4]. To explain some measurement event outcomes, the two-time, two-state approach refers to the future state of the known universe [483], which also seems of limited explanatory value [486, § 5].

These three approaches illustrate that introducing the concept of time symmetry into quantum mechanics is itself a challenge to satisfactorily understand, might generate further challenges to understanding, and might not reduce some of the existing challenges.

## 6.5 Spontaneous collapse theories: closed systems and conceptual anomalies

Spontaneous collapse theories focus on closed systems (see Sect. 3.1 above). These theories are intrinsically indeterministic: systems possess an irreducible disposition to spontaneously reach a definite value of position [487, 488] or energy [204]. Collapse theories use a modified Schrödinger equation so that the state suffers either an occasional hit leading to a spontaneous collapse [489], or is continuously driven towards such a collapse [490]. This can be seen as equivalent to the practical application of an unmodified Schrödinger equation in realistic situations [491]. The collapse involves the disappearance of the interference terms (see Sects. 3.5 and 3.6 above) in superposition and entangled states [106]. Collapse theories thus achieve similar results to those of quantum statistical mechanics (outlined in Sect. 4.3 above), but by a different method [270].

The details of the modifications leading to collapse are such that quantum mechanical states for simple systems follow Schrödinger evolution almost all the time, but states for large systems have frequent collapses [489, 490]. This account allows, in principle, the application of quantum mechanics to closed, unobserved systems (although there is a sense in which a collapse without an apparatus performing a measurement can only be hypothetical). It also, in practice, accounts for single measurement event outcomes. In collapse theories, probabilities relate to future events and interactions among physical systems, regardless of whether or not such events and interactions involve any notion of measurement [106, § 2]. When, in practice, collapse theories are applied to measurements, the apparatus is treated as any other system would be [487, § 6.5]. The predictions of spontaneous collapse theories match those of quantum mechanics for all experiments which are currently feasible. Some predictions of spontaneous collapse theories do diverge from those of quantum mechanics but, to date, no tests have been able to explore these limits [319, 492].

The state featuring in collapse theories appears to be informationally complete, in the sense that it is directly linked to physical independent reality. One approach takes the state as representing all there is [53, § 2]. The more common approach adds an additional ontology to the core principles of collapse theories [1, 228, 269], or at least recognizes that some such ontology is implied by the state [493, 494].

Either way, some challenges arise. With or without an added or implied ontology, collapse theories must link events to the mathematical state. Such links are almost always fuzzy: different physical events relate to arbitrarily similar mathematical states [495]. Fuzzy links imply that some probabilities prescribed by the mathematical state are ignored [62, § 4.3]. Such ignored probabilities are known as tails [496].

The question arises of how to treat an overlap of tails relating to states representing distinct physical systems. Depending on the treatment of such tails, it has been suggested that a range of anomalies can arise [62, § 4.3]. The counter-suggestion is that the anomalies arise only from an implied, inappropriate, understanding of probabilities prescribed by the state as relating to measurement events [487, § 11]. As noted above, probabilities in collapse theories relate only to future events and interactions among closed systems.

For those who accept the existence of anomalies, linking the state to a mass (or matter) density is claimed to avoid some of them [431, 497], but this is disputed [496]. An alternative takes only the collapses as representing the distribution of matter [498]: collapses are events at spacetime points and macroscopic objects are galaxies of such events [488][499, § 3]. This is in line with the possibility, noted in Sect. 2.1 above, that systems may be comprised of discrete events, rather than continually existing.

Even with anomalies, interpretations based on fuzzy or mass links can be maintained by modifying other assumptions [62, § 4.3][489, 495, 496]. Likewise, the event-based view is challenging but not necessarily incoherent [174, 500].

Collapse theories provide further evidence that a theory more descriptive of physical systems in isolation than is quantum mechanics is possible. Such theories also further illustrate the conceptual anomalies and challenges which can arise in linking a quantum mechanical formalism to independent reality.

## 6.6 Information-based approaches: insights on formalisms, few on reality

As noted in Sect. 2.1 and 3.2 above, the quantum mechanical formalism rarely describes individual physical systems, but generally prescribes only particular aspects of expected future events relating to physical systems. These features of the formalism, and similarities between it and theories of information and probability, suggest a range of possibilities.

- Quantum mechanics might reflect a limit on the amount of knowledge one can have about any system [112, 176, 210, 501–504].

- It might result from the constraints of algebraic and order-theoretic symmetries [505].
- It might reflect a limit on copying information [506].
- It might be such as to allow for maximal control of randomness [507].
- It might be a new type of probability theory [127, 508, 509].

Some suggest that quantum mechanics can be understood in terms of properties of information sources and communication channels in nature, rather than in terms of ontological features that give rise to these information-theoretic properties [506, 510].

Several questions remain unanswered by a purely information-based approach to understanding quantum mechanics [511, 512].

- What physically supports the information [176, § 3(a)][513, § 4]?
- How does this view apply to an apparatus [16, § 8][514, § 4.2]?
- How can quantum mechanics be assessed [515, § 6]?
- How can independent reality behave for quantum mechanics to be true [6, 35]?

One response to this last question is QBism [112], one form of quantum Bayesianism. In QBism, physical systems are real and independent of us and some features of the formalism (other than the state) may reveal aspects of independent reality [112, 171, 510, 516]. The state, however, is merely an expression of subjective information about the consequences of the quantum mechanics user's interventions into nature [516]. There is no true state [516, fn. i]. Quantum mechanical states are subjective [112, § III], one form of Bayesian probabilities (see Sect. 1.7 above): epistemic and personalist [112] rather than logical [103, § 4].

The question of how an information-theoretic based approach can be linked to independent reality has also prompted some other responses [216].

- 'Properties' in quantum mechanics can be viewed as descriptions of how one system influences others, rather than attributes possessed by the system [314, 505].
- An information-based principle theory may reflect an underlying constructive theory [216], of the type outlined in Sect. 6.7 below.
- Quantum information theory can be linked with a relational view, based on structural realism, of the type mentioned in Sect. 6.9 below.
- Quantum mechanics may be a theory which reflects ontic and epistemic three-way connections, be-

tween a user of the theory, a physical system, and the information available to the user [396, 517].
- A device-independent approach suggests that quantum mechanics is not about physical systems but, rather, about how the structure of language is constrained by physical independent reality [16].

Overall, information-based approaches have made significant progress in identifying and highlighting features which the quantum mechanical formalism shares with information theory. Such approaches have, to date, made far less progress on identifying what the formalism might indicate about the nature of (agent) independent reality.

## 6.7 Prequantum classical theories: a more descriptive theory may be possible

Classical theories are being developed which aim to more fully describe physical systems, as they exist independent of measurement. Most of these theories relate to a pre-quantum, or submicroscopic, level of independent reality, which is assumed to underlie phenomena dealt with by quantum mechanics, and which may be deterministic or indeterministic.

These theories illustrate that phenomena (system/apparatus interactions) which require quantum mechanical analysis are not necessarily inconsistent with systems (and apparatus) which are adequately described by classical physics.

1. In prequantum classical statistical field theory quantum statistics emerge from a stochastic field analyzed in a classical probability space [213, 345, 518]. All the predictions of quantum mechanics are reproduced, including those for an entangled system [518]. Other predictions, beyond those of quantum mechanics, are being experimentally explored [272, 519].
   Related developments include two similar models: one based on particles, rather than fields [520], and the other deterministic, rather than stochastic [521]. Not directly related but similar approaches include one based on the interaction of a signal wave with a carrier wave [522], and another based on a retro-causal constraining of a classical field [523].
2. The cellular automaton interpretation assumes a system of cells with classical deterministic evolution equations for each cell, which only depend on the data in the adjacent cells [159, p. 51].
   The classical laws represented by these equations give strong correlations over vast distances, prevent any action inconsistent with these, and so create a link between the resetting of a measuring device and widely-separated ontic states [159, § 5.7.3].

In this approach, sub-microscopic states are ontological, but quantum mechanical states are non-ontological templates to deal with microscopic phenomena. Due to the correlations, measurement event outcomes need be determined only for factual measurements, not for many counterfactual alternatives. Measurement confirms that the observable measured has (and had) an ontological state, and some other observables do (and did) not [159, § 4.2, § 5.4, § 5.5].

Prototype automata exist for quantum mechanics with interactions, and with composite systems [524].

3. Digital mechanics represents an alternative route to explore the possibility of a discrete, deterministic, prequantum process [525, 526].

4. Other approaches also assume a exact theory underlying quantum mechanics [342, 527–530]. Some of these [531, 532] stress the importance of the system-apparatus interaction (see Sect. 4.1 above).

5. A radically different approach has shown that many of the so-called quantum phenomena can be described, without quantization of matter and light, by considering both light and electrons as continuous classical fields [533].

## 6.8 Subensembles, quantum measures and alternative formalisms

Some further approaches arise out of specific mathematical analyses of quantum mechanics.

1. The subensemble based approach [23], limits physical interpretation of the formalism to apparatus readings. This approach [23, 193] is consistent with the analysis outlined in Parts 2 and 4 above. It uses uses quantum statistical mechanics to analyze the interaction process between A and S, as briefly outlined at the end of Sect. 4.3 above.

The approach restricts the extent to which the abstract probabilities in the formalism are linked to independent reality. These probabilities are interpreted as relative frequencies of runs, but only to the extent that they relate to subensembles characterized by specific outcomes. It can thus explain how a well-defined outcome emerges in a single run from a formalism which deals only with ensembles, and how different single runs, from one initial state, may have different outcomes.

The subensemble based approach, therefore, illustrates a way to link the formalism to independent reality, without the difficulties (see Sect. 3.4 above) of fully identifying the state with that reality.

It also highlights the value of the precise use of non-mathematical language (Sects. 1.1 and 1.3 above). For example it crucially depends on distinctions among different understandings of probability (Sect. 1.7 above) intended at each stage, and the distinction between mathematical objects and expected measurement event outcomes [23].

2. The quantum measure approach suggests that the concept of histories may be more fundamental than that of states [86]. This approach considers generalized stochastic processes, analyzed by reference to spacetime histories. This approach to quantum mechanics does not depend on the notion of measurement.

A central question in this approach is 'what corresponds to the physical world?' One suggestion is that the physical world is represented by an answer to every yes-no physical question that can be asked about the world, once the class of spacetime histories has been fixed [534].

3. Other mathematically based approaches arise from the use of alternative formalisms. These include approaches based on multiple trajectories [191, 535], category theory [536] and topos theory [79, 109][188, § 2, § 5].

## 6.9 Other useful frameworks: modal, relational and logical approaches

Three other approaches to the formalism can be helpful.

1. In modal interpretations [102, 537–540], the quantum mechanical state (a) refers to a single system which possesses physical properties at all times, and (b) represents what may be the case (modalities): which properties the system may possess.

In contrast, the value state represents what is the case: the physical properties the system possesses. Measurement events are ordinary physical interactions, definite outcomes are predicted, and outcomes reflect apparatus properties.

Conceptual challenges arise [263, § 3][537]: one modal approach involves ontological propensities

[541]; another proposes that the realm of possibility is as real as that of actuality [102][123, § 5].

Unlike the approach taken in this review, modal approaches relate the quantum state to a single system, rather than to an ensemble of system/apparatus interactions.

Bohmian and relative state approaches can be seen as types of modal interpretation [538].

2. Quantum mechanics is relational [176, 482, 502, 542, 543]. A state prescribes probabilities for interactions between a system and other systems, such as an apparatus (see Sect. 3.2 above). This is one reason to consider state and value as relational notions [103, § 4][176, 180, 482].

As noted above, Sect. 4.1, collective outcomes are not ascribed to systems, nor to apparatus, but to the totality [167, app. C]. This is analogous to structural realism, in which relations do not require relata having intrinsic identity [222, 269], but it is unclear precisely how quantum mechanics links to structural realism [544][545, § 4.2].

More generally, the implications of the relational view for independent reality are neither clear nor straightforward [546, 547]. Some relational views involve a sparse ontology of discrete events happening at interactions between what are assumed to be physical systems, rather than a fuller ontology, of permanent systems that have well defined properties at each moment of a continuous time [176].

Care is, however, needed in applying relational analysis to quantum mechanics. For example, being entangled is not a relational property [73, 244]. Being entangled is a feature of a composite system state, due to the composite being treated as one system (see Sect. 3.6 above). Reduced density matrices can represent subsystems of a composite system, but such matrices correspond to separate, non-entangled, states for the systems treated separately (see Sect. 3.7 above).

A relational approach to an entangled state would allow for three different accounts: one dealing with the interaction of the whole composite system and another system external to it, and the other two dealing, for each subsystem, with the interaction of that subsystem and another system interacting only with that subsystem [50, § 7.1.2].

3. Specific logical frameworks and analyses can

   – highlight features of the formalism [548],

   – contribute to understanding how that formalism relates to independent reality [481, 549], and

   – assess whether or not quantum mechanics necessarily requires new notions of truth [550–554].

Modal, relational and logical analyses provide useful tools to explore, separately or with other approaches, the characteristics of quantum mechanics.

# 7 Summary of main findings

Quantum mechanics in general prescribes only regularities among multiple measurement events. This suggests the possibility that events, rather than systems, may form the fundamental ontology of independent reality.

Statistical balance is a core (but unexplained) feature of quantum mechanics.

   – The collective response of an ensemble of identically prepared systems, to differing measurement types, is intricately balanced.

   – For example, the empirical data rarely justify a claim that each member of the ensemble was, pre-measurement, such that a definite value could be attributed to the measured property.

The following comprehensive characterization of a quantum mechanical state can be synthesized from elements of various analyses.

   – A state prescribes, in probability terms, aspects of expected events relating to an ensemble of systems, in a range of possible situations.

   – Such situations include the systems remaining closed, and also include the systems interacting with other systems (such as with a measuring apparatus, or a wider environment).

   – For measurement, the state prescribes probability distributions, which reflect the statistical balance in collective outcomes, both within ensembles, and among ensembles for differing measurement types.

This characterization can, in principle, apply to systems which either are not, or cannot be, observed. It highlights, however, that quantum mechanics deals with reality independent of human thought in only a very limited way, and does not necessarily describe intrinsic features of the physical systems to which it is applied.

This characterization also yields helpful perspectives on pure states, entanglement, and measurement.

– For a pure state, the relevant ensemble is such that any subensemble is represented by that same state.
– Entanglement is merely a more complex form of statistical balance, here relating to measurements of more than one observable, on different subsystems, for a composite system represented by a single state.
– This in turn, through decoherence theory, allows quantum mechanical analysis to extend to the environment of the measurement, and to closed systems.

Extrinsic decoherence theory shows how interference terms can become negligible. Quantum statistical mechanics shows how the other terms reach thermal equilibrium. This solves the measurement problem: it explains how a well-defined outcome may emerge in a single run from a formalism which deals only with ensembles.

There is no obvious way to formally apply the quantum mechanics uncertainty relations to an individual system, in terms of values assigned to observables, or properties possessed.

A theory more descriptive of independent reality than is quantum mechanics may yet be possible. The Bell and Bell-Kochen-Specker analyses do not necessarily rule out such theories, given the lack of consensus on their interpretation. There are many approaches to the pursuit of a more descriptive theory.

Bohr's approach dealt only with events. Bohmian approaches deal with systems and stress the fundamental role of the apparatus.

Significant challenges arise in any attempt to directly link the formalism to the ontic states of closed physical systems, as illustrated by relative state interpretations and spontaneous collapse theories.

Other approaches involve a less direct link.

– Time-symmetric approaches involve additional challenging concepts.
– Information-based approaches provide limited insight on independent reality.
– Prequantum approaches explore the scope for classical theories underlying the formalism.
– Further approaches arise directly from the mathematics of alternative formalisms.
– Modal, relational and logical frameworks offer useful analytic tools.

Physics might, in principle, be able to more fully describe independent reality. A first step in the pursuit of that possibility is to reach a greater degree of consensus, among both physicists and philosophers, on how to understand quantum mechanics. This review has

– concentrated on the conceptual, not the mathematical, aspects of understanding quantum mechanics,
– used non-mathematical language with precision, and
– aimed to reflect enough of the literature to be representative of the current state of understanding in quantum mechanics.

In these ways, this review contributes to achieving that greater consensus, and so to that pursuit.

# 8 Glossary: intended meanings for some non-mathematical terms

As noted in Sect. 1.1 above, understanding quantum mechanics is hard, because many pre-quantum mechanical, concepts and word meanings may need to be modified, and such understanding will be even harder if non-mathematical language is not used precisely.

This glossary therefore clarifies the intended meaning of some of the non-mathematical terms used in this review. For each of these terms, the glossary either specifies a non-mathematical meaning, or notes that no single meaning need be chosen. Terms in *italics* have their own glossary entries.

**Bohmian**   relating to the approach suggested by de Broglie and developed by Bohm; see Sect. 6.2

**causal**   generally defined relative to a specific theoretical model [555][556, § 3][557, § 4]; causal notions are not necessarily needed to achieve understanding [38]; even where they are invoked, there is no consensus on how causation is to be defined [327][558, § 2][559]

**coherence**   (for *waves*) the absence of spatial dispersion; (for *superposition states*) the existence of *interference terms* [234]; (for *entangled states*) *statistical balance* in collective *outcomes*, among *ensembles* for *measurements* of differing *observables* on multiple subsystems [235, § 2.3]: see Sects. 3.5 and 3.6

**completeness**   feature of a theory with terms which specify all the *properties* of a *system* [53]

**contextuality**   generally defined relative to one specific approach to *quantum mechanics* [328, § V.E][371]; broadly, assignment of values depends on the relevant *measurement* process [306]; see Sect. 5.7

**counterfactual**   (formal) a subjunctive conditional sentence, whose antecedent is contrary-to-fact [558]; (informal) relating to circumstances being other than as they are [30]; a complex area [327] **decoherence** absence of *coherence*; the disappearance of the *interference terms* in a *density matrix* representing a *superposition state*; see Sect. 4.4

**density matrix** a matrix, with specific mathematical characteristics, over a *Hilbert space*

**determinism** implies that any possible group of *systems* (which are isolated from any other *systems*) will evolve in a single unique way from any possible initial conditions [138]; *ontological* determinism implies that determinism is a feature of *independent reality*; *epistemic* determinism implies that determinism is a feature of our knowledge of *independent reality*; see Sect. 1.9

**empirical** relating to *phenomena* collectively

**entangled state** *superposition* of composite *system pure states*; implies *statistical balance* in collective *outcomes*, among *ensembles* for *measurements* of differing *observables* on different subsystems; see Sect. 3.6

**ensemble** *statistical ensemble*

**epistemic** refers to the ways in which humans acquire knowledge and process information [25, p. 57][233]

**ergodicity** the coincidence of time and *ensemble* averages in *empirical* data [337, 338]

**event** several meanings are possible [174, § 2]; this review will take event to mean the instantiation of one or more *properties* within some region of spacetime

**field** specification of *properties* smoothly across spacetime [25, p. 57]

**formalism** one of several mathematical structures, in which analysis in *quantum mechanics* can be carried out; examples include *Hilbert spaces*, Fock spaces [560], quasi-sets [560], Dirac brackets [277, § 5], Feynman path amplitudes [128, § 1], multiple trajectories [191, 535] and algebraic theories [561] such as category theory [536] and topos theory [79, 109][188, § 2, § 5].

**hidden variables** variables, other than those specified by the *quantum mechanical formalism*, which some approaches to *quantum mechanics* assume are needed to entirely specify the *ontic state*; see Sect. 6.2

**Hilbert space** a mathematical vector space, usually with many more than 3 dimensions; one of the *formalisms* [25, ch. 2]

**incompleteness** lack of *completeness*

**independent reality** that which exists other than only in human thought; see Sect. 1.5

**interference** the combined effect of the effects of several *waves*; typically features regions where these effects cancel each other [24, § 3, p. 131]

**interference terms** in a *superposition state*, terms representing, not *measurement event outcomes* but, rather, *statistical balance* in collective *outcomes*, among *ensembles* for differing *measurement* types; see Sect. 3.5

**intrinsic decoherence** disappearance of *interference terms* in a coarse-grained approximate mathematical analysis of a closed *system* (notionally) split into an open subsystem and an ignored residual subsystem; see Sect. 4.4

**locality** no single definition; usually implies some limitation on the extent to which one *system* or *measurement event* can influence, link to, or affect, another spatially separated *system* or *measurement event*; see Sects. 5.4 to 5.6; occasionally may be given a temporal meaning [88]

**matter** *particles*, *waves* or *fields* singly or collectively

**measurement** dynamical process in which apparatus A is repeatedly coupled to successive members, S, of an *ensemble* of *systems* to explore all possible values of a joint *property* of S and A; the *result* is ascribed to the whole closed *phenomenon*; see Sect. 4.1

**measurement event** *single run* of a *measurement*

**mind** no single definition is assumed

**mixed state** a *state* for which the relevant *ensemble* can be split, such that each subensemble is represented by a different *pure state*; see Sect. 3.3

**multi-field** specifies properties, smoothly across spacetime, by reference to *multiple* points in spacetime [230, 231]; see Sect. 3.4

**object** that which is being studied, discussed or examined

**objective** relating to an *independent reality* [49]; what is objective should not depend on the particular perspective used for the description [63, § 3.1]

**observables** *operators* associated with *properties* relating to the studied *system* [193, 300]

**observe** to use physics concepts to account for what is done or thought about a *measurement*; to describe a *measurement* by associating data [562, 563]

**ontic state** a complete specification of the *properties* of a *system* (in an *ontological* model) [212]; see Sect. 3.4

**ontology/ontological** structures postulated in a physical theory as primary, underlying, explanatory entities [78, 564] to account for the existence of *events* [500]; can be seen as forming the basis for kinematics (all possible values and arrangements of the physical ontology) and dynamics (specific constraints on how the ontology evolves in time) [430]; alternatively can be based on the dynamics [494, § 2]; often expressed in terms of the nature and behaviour of *systems* as they are [564], independent of any *empirical* access [233]

**operator** map associating every vector in a *Hilbert space* with another such vector [25, p. 37]

**outcome** apparatus reading for a *single run* of a *measurement*; collectively form the *result* of a *measurement*

**particle**    entity associated with a group of spatially localized *properties*, some of which are unchanging [25, § 4.2]; possibly has *self-identity* [17, 388, 565]

**phenomenon**    *observation* linked to specified circumstances, including (where appropriate) the experimental set-up [404, p. 64][405, § 3.3]

**prepare/preparation**    (of a *system*): *selection* of some of the *single runs* of a *measurement* [193]

**pre-quantum mechanical**    developed prior to the experimental exploration of subatomic *phenomena*

**prequantum**    relating to a level of *independent reality* which is assumed to underlie the *phenomena* dealt with by *quantum mechanics*; see Sect. 6.7

**prescribe**    to specify (or, literally, write) in advance

**property**    the values of some specified class of physical quantities lying in specified ranges [25, § 4.2]; may relate to a *system* considered in isolation but, in *quantum mechanics*, usually relates jointly to the *system* and either an apparatus or a wider environment [563, 566]

**pure state**    a *state* for which the relevant *ensemble* is such that any subensemble of that *ensemble* is represented by the same *state*; see Sect. 3.3

**quanta**    *particle*-like concept in *quantum theory*, particularly *quantum field theory*; see Sect. 5.8

**quantum field theory**    *quantum theory* dealing with *phenomena* described in the integrated spacetime of special relativity; see Sect. 1.6

**quantum mechanics**    one or more of the *formalisms* of *quantum theory*, excluding any parts of those *formalisms* specific to *quantum field theory*

**quantum theory**    group of theories developed in response to the exploration of subatomic *phenomena*, but, in this review, excluding the theories, mentioned in Sect. 1.6, which seek to incorporate general relativity.

**realism**    any view which considers the notion of *independent reality* meaningful; see Sect. 1.5

**reduced density matrix**    *density matrix* related to a subsystem of a composite *system*; gives probabilities for *outcomes* of *measurements* restricted to the subsystem; not a *state* adequately representing the subsystem in the context of the wider composite system, but a coarse-graining of the composite *state*; see Sect. 3.7

**relative state**    the approach outlined in Sect. 6.3

**result**    collective *outcomes* of a *measurement*

**Schrödinger equation**    an equation specifying how the quantum *state* evolves in time [25, § 6.4]

**selection**    using *single run outcomes* to split an *ensemble* into subensembles, each with one value of the relevant *observable* [193]

**self-identity**    feature of an entity for which different instances of it can be distinguished, like currency coins

(and unlike currency in a bank account) [567, § 8]; the extent to which *quanta* have self-identity is unresolved [17][56, ch. 3][374, 376, 384, 388, 565, 568]

**single run**    (in *measurement*): one *system* (from an *ensemble*) interacting with an apparatus, followed by reading of an *outcome* [193]; see Sect. 4.1

**state**    mathematical term which *prescribes*, generally in terms of probability distributions, aspects of expected future events relating to a *statistical ensemble* of *systems* in a range of possible situations; see Sect. 3.2

**statistical balance**    sense in which, for some combinations of *measurement* types, the collective responses, of a *statistical ensemble* of *systems*, to differing *measurement* types, is intricately balanced, as reflected in *empirical* data confirming probabilities *prescribed* by *quantum mechanics*; see Sect. 2.2

**statistical ensemble**    a set of *systems* which can be treated as identical, such as those *prepared* in an identical way [193]

**subensemble**    part of a *statistical ensemble* which is itself a *statistical ensemble* [193]

**superposition**    a mathematical combination of *pure states* to form another *pure state*; see Sect. 3.5

**system**    an *object* which can be isolated well enough, and specified clearly enough, to allow it to be studied [25, 566]; in this review, system is used extensively as an alternative to *wave*, *particle*, *field* or *object*, to avoid connotations of such concepts in classical physics or general intuition, which may not be appropriate in *quantum mechanics*; care is still needed to avoid some remaining unhelpful connotations, and limitations [569], of the word system itself [16]; likewise intuitive connotations of subsystem may also be unhelpful [570]; systems may be thought of as comprised of discrete *events*, rather than having a continuous existence [174, 176]

**wave**    a *field* which evolves in time and exhibits characteristic behaviour such as *interference*

**Acknowledgements:** I acknowledge the work of all the authors of the cited references and, in particular, those with whom I have spoken. Their work has made mine possible. I also thank the editors and anonymous reviewers for their insight and constructive comments on earlier versions. Their work has made mine better.